**Revealing the $MoS_2$ Growth Mechanism in Chemical Vapor Deposition: Real-Time Imaging and Statistical Analysis**

Faizal Arifurrahman[1,2†], Tianshu Zhai[1†], Yuguo Wang[1], Jing Zhang[1], Andrew L. Hitt[1], Jun Lou[1,3*], Ming Tang[1,3*]

1. Department of Materials Science and Nanoengineering, Rice University, Houston, TX

2. Faculty of Mechanical and Aerospace Engineering, Institut Teknologi Bandung, Jalan Ganesha 10, Bandung, 40132, Indonesia

3. Rice Advanced Materials Institute, Rice University, Houston, TX 77005, USA

[†] These authors contributed equally.

[*] Corresponding author emails: jlou@rice.edu, mingtang@rice.edu

**Abstract**

Chemical Vapor Deposition (CVD) is a promising method for scalable synthesis of two-dimensional transitional metal dichalcogenides (TMDs) such as $MoS_2$, but challenges in reproducibility and controllability persist due to an incomplete understanding of their dynamic growth mechanisms. While in-situ characterization methods could provide valuable insights, it remains challenging to track a large ensemble of crystals to enable quantitative, statistical analysis. Here, we address this gap by developing and applying a semi-automated image processing pipeline to analyze in-situ optical microscopy footage of $MoS_2$ growth. This framework enables the high-throughput reconstruction of complete growth trajectories for over 400 individual crystals from a single experiment. Our statistical analysis demonstrates that $MoS_2$ crystallization is governed by an edge-attachment-limited mechanism rather than by precursor diffusion. Furthermore, $MoS_2$ crystals exhibit non-competitive growth, indicating that precursor supply does not limit the growth of neighboring flakes until physical impingement occurs. These findings provide direct, quantitative evidence that advances the fundamental understanding of TMD growth, establishing a powerful methodology for rational optimization of the CVD growth of two-dimensional materials.

## Introduction

Two-dimensional (2D) materials including graphene,[1–3] hexagonal boron nitride[4–7] and transition metal dichalcogenides (TMDs)[8–11], have been the subject of intensive research over the past decade, revealing a wealth of exceptional properties from high carrier mobility to layer-dependent bandgaps. A primary challenge in harnessing these properties for next-generation device applications such as field-effect transistors, gas sensors, and photodetectors lies in developing scalable synthesis methods that produce high-quality, large-area 2D materials with well-controlled defect population and distributions.[12–23] While mechanical exfoliation can produce large flakes for fundamental studies,[24–27] chemical vapor deposition (CVD) has emerged as one of the most promising fabrication techniques for mass manufacturing.[28–31] Despite significant progress, CVD growth of 2D materials continues to face challenges in reproducibility and controllability. Subtle variations in growth conditions such as reactor geometry, precursor ratios and temperature profiles often lead to inconsistent crystal morphology and defect population, hindering the transition of 2D materials from the laboratory to industrial applications.[32] These challenges stem from an incomplete knowledge of the 2D crystal growth mechanisms during CVD processes.

To establish a more thorough mechanistic understanding and guide the optimization of CVD processes, in-situ characterization techniques are invaluable. Direct observation of crystal nucleation and growth can provide critical insights into transient phenomena such as crystal nucleation and intermediate phases that are often missed in conventional ex-situ analyses, which only capture snapshots of the initial and final states. Recent advances in real-time monitoring unveil distinct growth modes of $MoS_2$ due to changes in the gasification behavior of $MoO_3$ precursors,[33] growth and self-assembly of graphene flakes on liquid metal,[34] and the dynamic growth process of 2D van der Waals heterostructures (e.g. $CdI_2/MoS_2$, $PbI_2/WS_2$).[35] While high-magnification characterization methods like in-situ transmission electron microscopy (TEM) can reveal the early growth stages,[36,37] their narrow field of view restricts the number of crystals that could be studied and the special reaction chambers required may not fully represent typical CVD settings. In comparison, in-situ optical microscopy could capture the evolution of many crystals within a large region and also under more realistic synthesis conditions.[38–41] However, existing optical imaging studies have generally been limited to qualitative observation or a small number of crystals. A comprehensive, quantitative analysis of a large 2D crystal population to yield statistical insights has remained elusive, largely because manually tracking the shape evolution of individual crystals is very labor-intensive.[42–46]

In a previous work, we integrated a miniaturized CVD apparatus with optical microscopy and image-processing capabilities.[8] This system was used to acquire real-time footage of monolayer $MoS_2$ crystal growth, and the resultant observations were leveraged to guide the synthesis process and the manipulation of crystal morphologies. The study successfully demonstrates an effective workflow that combines in-situ observation with image data analysis and machine learning predictions to achieve 2D crystal growth with controlled properties. However, one important aspect missing from that work is the extraction and analysis of the growth trajectories of individual $MoS_2$ crystals, which should facilitate a deeper understanding of the fundamental growth mechanisms.

Building upon our prior study, here we introduce a semi-automated image processing pipeline designed to systematically track the nucleation and growth of a large ensemble of $MoS_2$ flakes from in-situ optical videos. By applying this framework, the growth history of over 474 individual crystals was extracted from a single experiment, enabling a robust statistical investigation of their behavior. We comprehensively examined the growth rates of these $MoS_2$ crystals and their correlation with crystal size, orientation and surrounding crystal distribution. Our analysis reveals that the $MoS_2$ growth kinetics are primarily controlled by the precursor attachment to crystal edges rather than mass transport. Furthermore, we observed that neighbor $MoS_2$ crystals do not compete for precursors and exhibit independent growth until their coalescence. Such behavior is attributed to the presence of $MoO_xS_y$ nanoparticles at the center of $MoS_2$ flakes, which supply Mo feedstocks to crystal growth from within. These insights advance our understanding of the CVD synthesis of $MoS_2$ and help pave the way towards more reproducible synthesis of 2D materials.

**Results & Discussion**

A custom-designed miniaturized CVD reactor, featuring a rectangular quartz tube, was employed to synthesize $MoS_2$ on a c-plane (0001) sapphire ($Al_2O_3$) substrate as shown in **Error! Reference source not found.** Real-time monitoring of $MoS_2$ growth was achieved by placing an objective lens with an ultra-long working distance above the substrate, which is connected to a digital camera. During a typical growth experiment conducted within this setup, where the reactor temperature was ramped from room temperature to 890 °C at a rate of 100 °C/min, the recorded video footage shows that triangular $MoS_2$ flakes emerged, expanded and merged across the substrate, see snapshots in **Error! Reference source not found.** and Supplementary Video 1 in Supplementary Information. **Figure 2** presents the corresponding segmented images at 0, 30, 60, and 150 s after the first nucleation of $MoS_2$ was observed, illustrating the temporal evolution of the $MoS_2$ crystal domains.

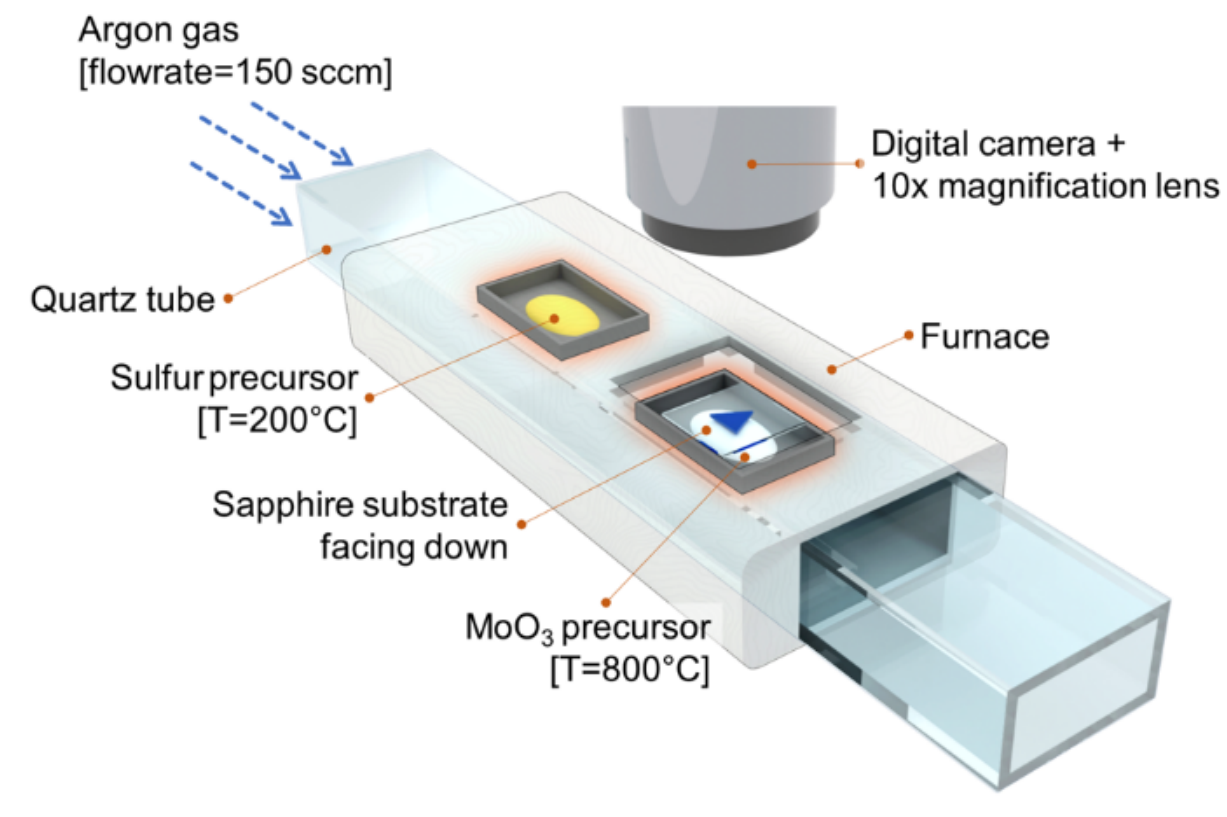

**Figure 1.** Schematics of the miniaturized CVD reactor.

Post-growth characterization confirmed the monolayer nature of the as-grown $MoS_2$ flakes. Raman spectroscopy revealed two distinctive peaks at 385.6 and 404.4 $cm^{-1}$, corresponding to the $E_{12g}$ and $A_{1g}$ vibration modes of $MoS_2$, respectively (Figure S2a). The peak separation of 18.8 $cm^{-1}$ is characteristic of $MoS_2$ monolayers, which is consistent with a strong excitonic peak at ~660 nm in the photoluminescence spectrum (Figure S2b). Atomic force microscopy (AFM) further corroborated this, showing a uniform height of 0.6 nm across the crystals, matching the expected thickness of monolayer $MoS_2$ (Figure S3).

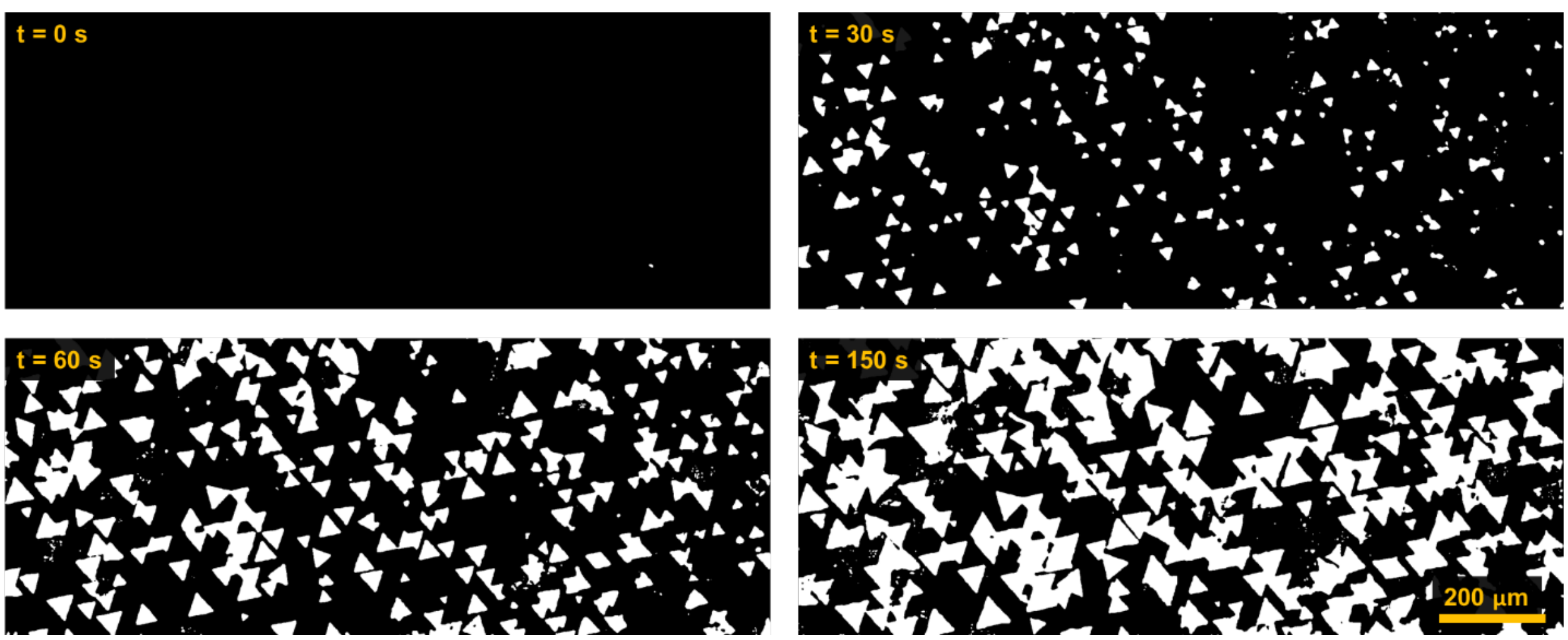

**Figure 2**. Snapshots of an in-situ optical video showing the formation and evolution of triangular $MoS_2$ crystals on the (0001) sapphire substrate at 0, 30, 60, and 150 seconds after first $MoS_2$ nucleation events.

The in-situ optical video footage provides a detailed view of the $MoS_2$ growth process. In the experiment, burst nucleation of $MoS_2$ was observed when the reactor temperature reached c.a. 860°C, where a large number of crystals suddenly appeared and began rapidly growing on the substrate. As shown in **Figure 3**a, the nucleation rate remained roughly constant during the initial phase of ~20 seconds, and the total crystal area increased with time $t$ as $A_{tot} \propto t^{1.8}$, with $t = 0$ s defined as the moment when the first crystal was detected. Notably, after $t \approx 20$ seconds, both nucleation and growth of $MoS_2$ began to slow down significantly. By $t \approx 100$ s, crystal growth effectively stagnated, with minimal increase in $MoS_2$ coverage afterwards. This deceleration suggests a continuous reduction in the thermodynamic driving force during the CVD process. Given that excess sulfur was found to remain in the reactor after the experiment ended, we attribute this primarily to the depletion of Mo precursors rather than sulfur species.

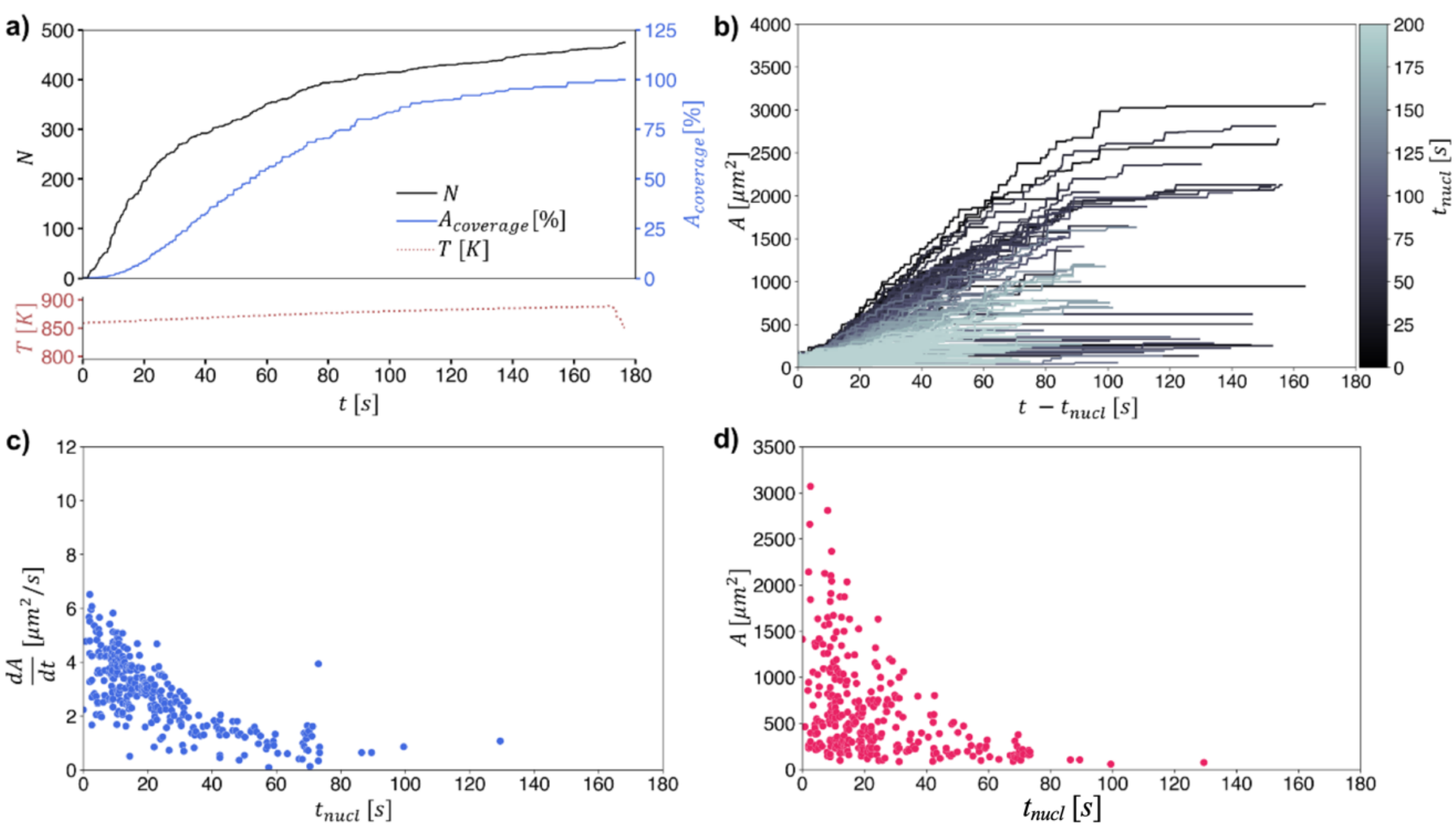

**Figure 3.** $MoS_2$ nucleation and growth dynamics revealed by in-situ optical observation. (a) Number and relative area coverage of $MoS_2$ flakes (top) and reactor temperature profile (bottom). (b) Area of individual $MoS_2$ crystals versus growth time defined as $t$ - $t_{nucl}$, where $t_{nucl}$ is the time a crystal was first detected. The curve color reflects $t_{nucl}$. (c) $MoS_2$ growth rate vs $t_{nucl}$. (d) Final $MoS_2$ crystal size, which is measured either at the end of the experiment or when the crystal coalesced with other $MoS_2$ flakes, vs $t_{nucl}$.

***Growth kinetics of individual crystals***

By applying the semi-automated processing workflow to the in-situ optical video, we were able to track the growth history of each individual $MoS_2$ crystal that appears in the footage. **Figure 3**b presents the size evolution of 474 crystals over time. To facilitate comparison, their growth curves are shifted horizontally so that they all start from the origin, which corresponds to the time a crystal was first detected. The growth curve of a $MoS_2$ crystal is terminated on the plot either when the end of the experiment was reached or the crystal merged with a neighbor. While these crystals display a wide range of growth rates and final sizes, their growth trajectories commonly display two distinct stages, i.e. a "steady-state" period where the crystal size increased approximately linearly with time, followed by a stagnation period where growth effectively ceased. Some crystals did not exhibit the second stage in **Figure 3**b because they coalesced with other crystals before growth slowed down. The individual growth trajectories are consistent with the ensemble behavior shown in **Figure 3**a.

To quantitatively analyze the growth process, we plotted the steady-state growth rate and the final size of all tracked $MoS_2$ flakes against their nominal nucleation time (i.e. the time they were first detected), in **Figure 3**c and 3d, respectively. The growth rate shown in **Figure 3**c is calculated from the average slopes of the quasi-linear portion of the growth curves. These plots reveal a clear trend: crystals that nucleated earlier grew faster and also attained larger sizes. The $MoS_2$ growth speed decayed rapidly with increasing nucleation time, reflecting the progressive decline in precursor supersaturation within the reactor. Later-nucleating crystals were smaller because they grew more slowly and also had less time to grow. We note that many crystals that nucleated in the early stage also have small apparent sizes because they quickly merged with other crystals after nucleation.

***Diffusion- vs. Edge-attachment-limited growth***

The CVD growth of 2D materials, such as graphene, is known to proceed via either diffusion-limited or edge-attachment-limited mechanisms, depending on experimental conditions such as precursor supersaturation.[47,48] In the diffusion-limited regime, crystal growth is rate-controlled by the transport of precursor species from the surrounding to the crystal edges. Crystal size typically increases at a constant speed in such a growth mode[49]:

$$\frac{dA}{dt} = \text{constant} \tag{1}$$

Conversely, edge-attachment-limited growth occurs when precursor diffusion is facile relative to their incorporation into the lattice. Crystal growth rate in this mode exhibits a parabolic size dependence:

$$\frac{dA}{dt} \propto \sqrt{A} \tag{2}$$

Compared to graphene, whether TMD growth is diffusion- or edge-attachment-limited is less clear despite several modeling studies.[50–54]

Direct measurement of individual $MoS_2$ crystal growth rates from the optical video enables us to distinguish the dominant growth mechanism. It should be pointed out that the quasi-linear growth ($A \propto t$) exhibited by individual $MoS_2$ is not a sure sign of a diffusion-limited mechanism. That is because it may also result from edge-attachment-limited growth with a decaying edge velocity due to precursor depletion. To resolve this ambiguity, we compared the instantaneous growth rates ($dA/dt$) of different crystals at the same time during growth, since all crystals should experience similar supersaturation. As shown in **Figure 4**, we calculated $dA/dt$ at $t$ = 50 seconds after the detection of the first $MoS_2$ flake based on the linear regression of the $MoS_2$ growth curves between 46 s and 54 s. It exhibits a clear positive correlation with crystal size, i.e. larger crystals also grew faster. This trend is inconsistent with diffusion-limited growth, where $dA/dt$ should be size-independent, but aligns with edge-attachment-limited kinetics. In this regime, larger crystals have larger $dA/dt$ because their longer edges provide more sites for precursor incorporation. Such size dependence of the instantaneous growth rate was also evident at other times (see Figure S4). This result confirms that $MoS_2$ growth is limited by interface reaction kinetics at relatively high reaction temperatures above 850 °C, with precursor transport remaining facile over a wide range of supersaturations as experienced in the experiments.

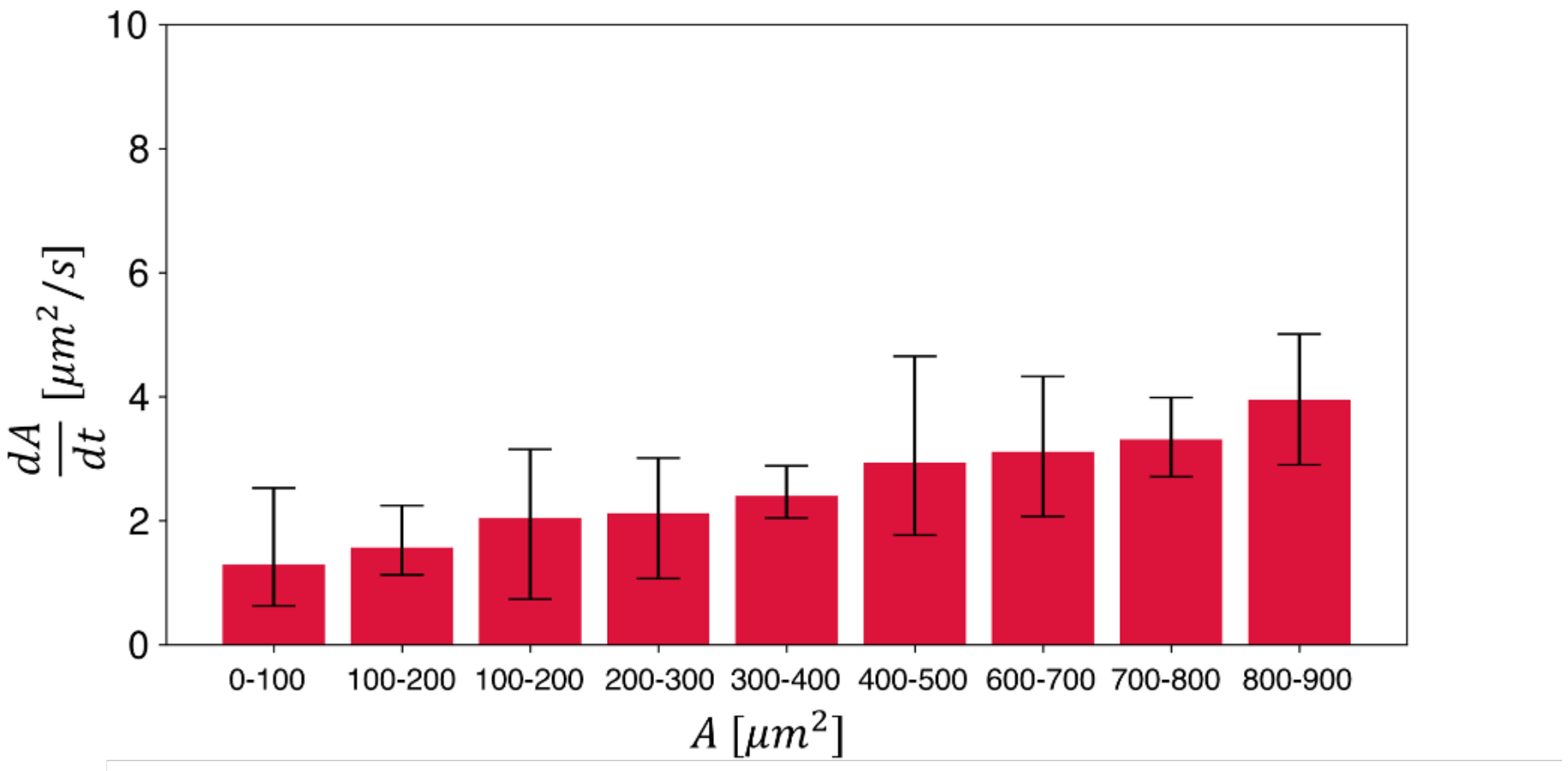

**Figure 4.** Instantaneous $MoS_2$ growth rate $dA/dt$ versus crystal area at 50 s after the first $MoS_2$ was detected.

In contrast to the detailed single crystal analysis performed here, the classic Johnson-Mehl-Avrami-Kolmogolov (JMAK) model[55–57] is a tool widely used to analyze phase transformation kinetics under isothermal condition when only aggregated information of the transformation process is available. This involves fitting the experimental data to the JMAK equation

$$f = 1 - \exp(-kt^n) \qquad 3)$$

where $f$ is the transformed fraction of the system, and $n$ and $k$ are parameters known as the Avrami exponent and the rate constant $k$, respectively. The Avrami exponent is often expressed as

$$n = a + bc \qquad 4)$$

where $a$ is the nucleation index, $b$ is the growth dimensionality, and $c$ equals 1 or 1/2 for interface- or diffusion-limited growth, respectively. $a = 0$ corresponds to instantaneous nucleation, $a = 1$ for constant-rate nucleation, and $0 < a < 1$ represents continuous nucleation with a decaying rate. Finding a reasonable set of values for ($a$, $b$, $c$) helps interpret the nucleation and growth behavior. Fitting the area coverage of $MoS_2$ vs time curve (Figure 3a) with Eq. 3 yields an Avrami exponent of 1.71. The JMAK analysis alone may prompt the assignment of $a = 0.71$, $b = 2$ and $c = ½$, which leads to the incorrect conclusion that $MoS_2$ growth is diffusion-limited. This demonstrates the limitation of the traditional JMAK analysis and highlights the value of in-situ imaging of individual crystals coupled with statistical analysis in obtaining an accurate understanding of 2D material synthesis.

***Effect of $MoS_2$/substrate orientation relation***

Due to van der Waals epitaxy, TMD crystals grown on c-plane sapphire substrate typically show preferential alignment along specific crystallographic directions. The $MoS_2/Al_2O_3$ interaction energy is minimized when $\langle 10\bar{1}0 \rangle$ of $\alpha$-$Al_2O_3$ is parallel to the $\langle 10\bar{1}0 \rangle$ or $\langle 01\bar{1}0 \rangle$ zigzag edges of $MoS_2$, which are denoted as the 0° / 60° alignments, respectively.[29] Metastable $\pm$ 30° misorientations between $MoS_2$ and $Al_2O_3$ have also been observed, which represent the local energy minima.[58] Our in-situ optical video revealed two predominant orientation groups separated by 60°, which should correspond to the preferred 0°/60° alignments. Because optical microscopy alone cannot distinguish the $MoS_2$/sapphire orientation relation, one group was somewhat arbitrarily assigned to 0° and the other to 60°. To quantify the orientation distribution of $MoS_2$ crystals, we measured the orientations of all resolvable crystals by fitting regular triangles to their segmented images either at the end of the video footage or right before they merged with neighbors. Small crystals (<10 $\mu m^2$) were excluded from the analysis due to limited image resolution. When assigning crystals to one of the two major orientations (0° and 60°), a ±15° tolerance was used to account for measurement uncertainties from instrument vibration, air flow fluctuations, and segmentation artifacts, etc. Figure S5 shows that crystals with 0° or 60° orientations were uniformly distributed across the substrate. Nevertheless, there are more 0°-oriented (50% of all crystals) than 60°-oriented (36%) crystals, see Figure 5a. This asymmetry may arise from factors such as the substrate conditions (e.g., miscut angle between the substrate surface and (0001) plane[59]) and carrier gas flow.

The existence of preferred $MoS_2$-substrate alignment prompted us to investigate whether the crystal growth rate is orientation-dependent. Figure 5b shows that the growth speeds of crystals with 0°, 60° and other orientations are not statistically distinguishable. This suggests that while the crystal-substrate interaction strongly influences nucleation rate by favoring certain epitaxy relations, it has minimal impact on the edge attachment kinetics, rendering the growth rate insensitive to crystal orientation.

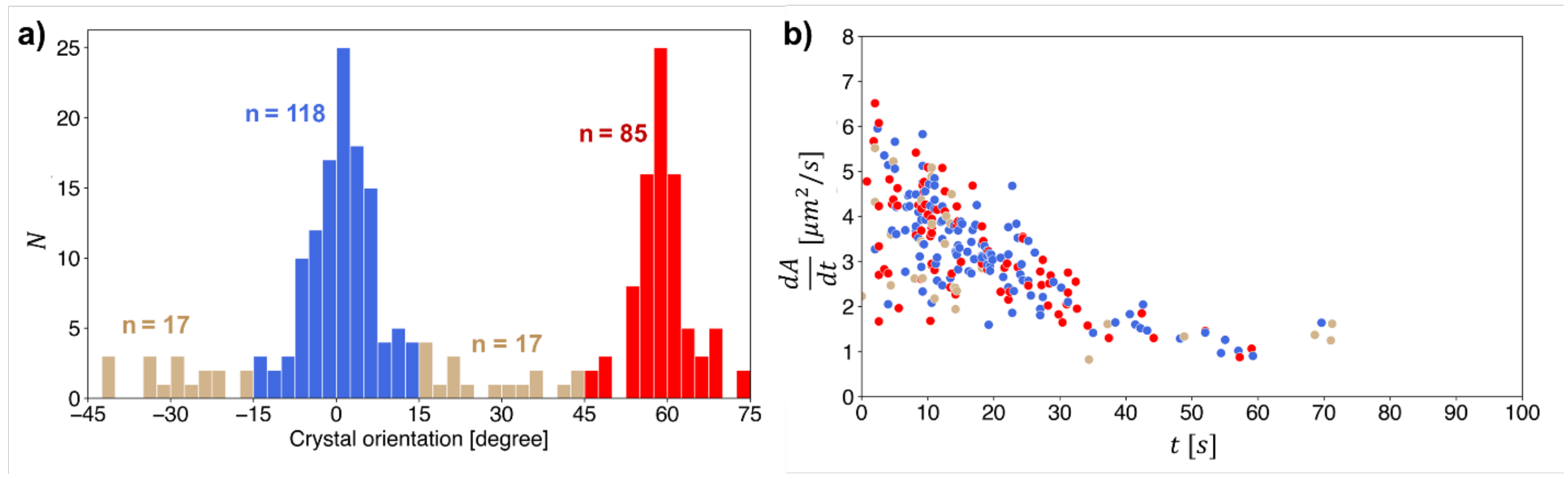

**Figure 5.** (a) Orientation distribution of $MoS_2$ crystals on the c-plane sapphire substrate. The blue and red groups represent the two preferred orientations (0° and 60°) with a 15° tolerance. (b) Comparison of the growth rates of $MoS_2$ with different orientations: blue – 0°, red – 60°, and tan – other.

***Growth competition between $MoS_2$ crystals***

During crystal growth, it is common that neighboring crystals compete for feedstocks when their diffusion fields overlap, leading to phenomena like nucleation/growth depletion zone.[60] We examined whether such competition is present during $MoS_2$ growth and affects the overall growth kinetics in our experiment. In Figure 6a and 6b, individual $MoS_2$ flakes at $t$ = 50 s after the detection of the first crystal were divided into different groups based on the number ($N_{surr}$) of their neighbor crystals. A crystal is considered a neighbor if the centroid distance is less than $3R$, where $R = \sqrt{A/\pi}$ is the equivalent crystal radius. In Figure 6b, the flaks are grouped based on the area coverage ($f_{surr}$) of neighbor crystals within a radius of $3R$ from the crystal's centroid. These plots clearly show that the average crystal growth rate does not exhibit a noticeable correlation with $N_{surr}$ or $f_{surr}$, indicating that the local crystal density did not affect growth kinetics. The absence of competition between neighbor crystals persisted even at later times when precursor depletion was more severe (Figure S6). To further validate this conclusion, we tracked the morphological evolution of two $MoS_2$ crystals with parallel opposing edges upon approaching each other (Figure 6c). As illustrated in **Figure 6**d, if two neighbor crystals compete for precursors during growth, their facing edges would grow slower than other edges due to local precursor depletion, which will cause their centroids to move away from each other. However, such behavior is not observed in the experiment: **Figure 6**e shows that the centroid distance between the two crystals remained constant until they impinged against each other. For completeness, the centroid distances of crystal pairs with other orientation relations (0°/0°, 0°/30°) were also examined (Figure S7), which do not exhibit any significant variation over time either. These results provide convincing evidence that individual $MoS_2$ flakes grew independently and did not interfere with one another. This is in sharp contrast to the growth of neighboring graphene domains on the catalyst substrate, which experiences a gradual slowdown when they approach each other.[47] Different from the centroid distance, Figure S7g-i shows that the gap distance between the facing edges or vertices of two $MoS_2$ crystals is reduced at an increasingly slow speed, which was caused by the global decline of precursor supersaturation.

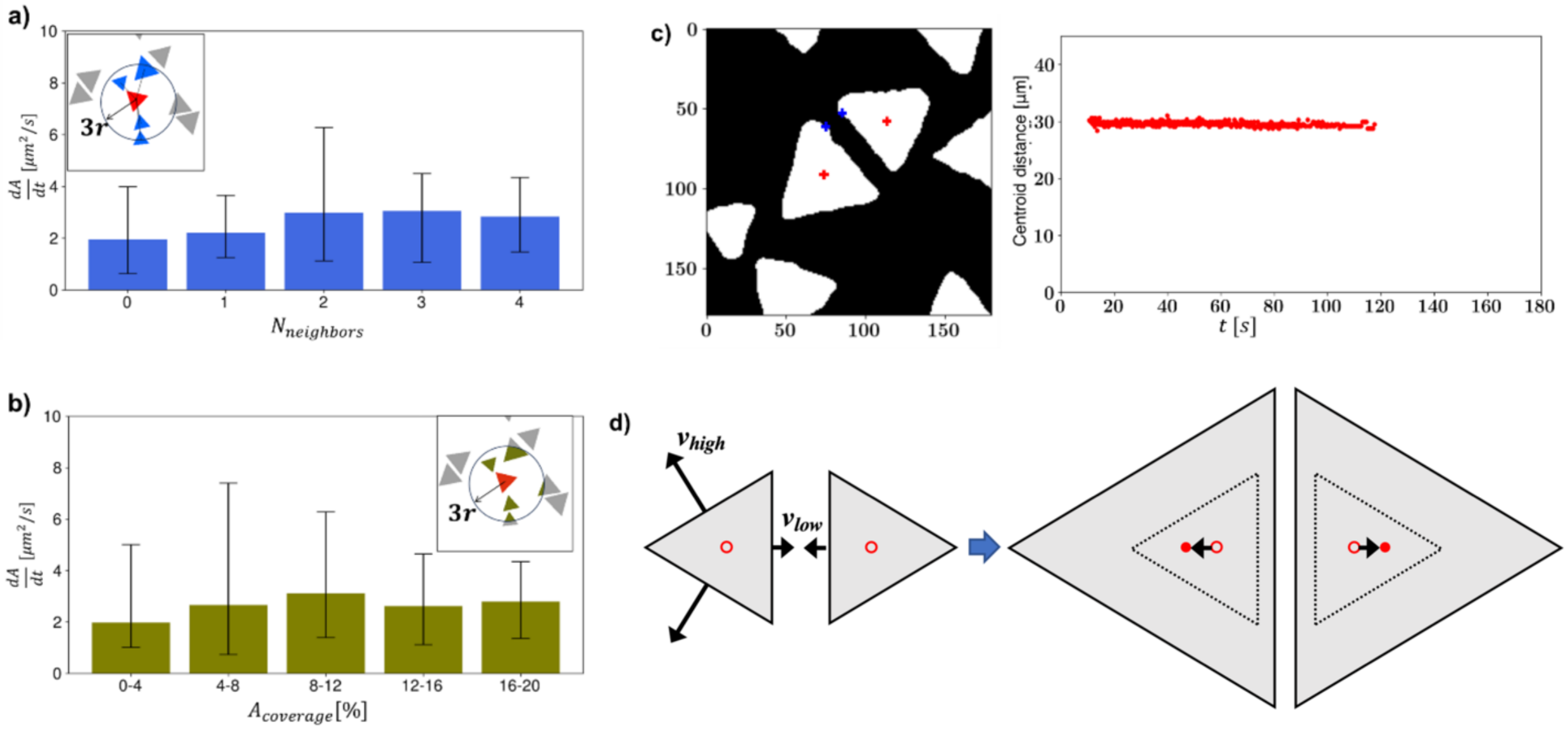

**Figure 6.** Analysis of $MoS_2$ crystal interaction during CVD growth. (a) Instantaneous crystal growth rate vs. number of neighbor flakes within a distance of 3*R* to the crystal. (b) Instantaneous crystal growth rate vs. area coverage of neighbor flakes within a distance of 3*R* to the crystal. (c) Two neighbor $MoS_2$ crystals with parallel opposing edges from the in-situ video (left) and their centroid distance as a function of time. (d) Hypothetical evolution of the centroid locations of two growing $MoS_2$ crystals if they compete for precursors.

### *Seeded $MoS_2$ crystal growth*

To better understand the origin of the absence of inter-crystal competition during $MoS_2$ growth, we examined the as-grown samples using scanning electron microscopy (SEM). **Figure 7**a reveals that nearly all triangular $MoS_2$ crystals contain a nanoparticle at their centers. The formation of these nanoparticles could be attributed to the relatively high Mo precursor temperature (850 °C) employed in our experiments. Previous work has shown that raising the Mo precursor temperature above 750 °C promote the deposition of molybdenum oxynitride ($MoO_xS_y$) nanoparticles onto substrates due to the increased vapor-phase supersaturation.[61,62] They were found to act as seeding centers, initiating the growth of $MoS_2$ flakes around them.[37,63]

We propose that the presence of $MoO_xS_y$ seeds is responsible for the non-competitive $MoS_2$ growth as observed in our study. As illustrated in Figure 7b, when a $MoO_xS_y$ nanoparticle is deposited onto the substrate, it serves as the localized Mo reservoir to continuously supply Mo feedstocks to the surrounding $MoS_2$ crystal. Because of their large diffusion coefficient on TMD surfaces[64], Mo feedstocks can diffuse rapidly across the $MoS_2$ terrace to reach crystal edges, where they are incorporated into the $MoS_2$ lattice. However, a Schwoebel-Ehrlich barrier[65,66] present at step edges, along with a much lower diffusivity on the sapphire substrate[64], prevents Mo feedstocks from escaping to the surrounding, effectively isolating each crystal's Mo supply. As each $MoS_2$ crystal draws Mo feedstocks from its own seed, adjacent crystals grow independently until physical impingement occurs. When Mo is primarily supplied by $MoO_xS_y$ seeds rather than from the vapor phase, the number of $MoS_2$ crystals formed on the substrate is controlled by the seed density. The confinement of Mo feedstocks within the seed-containing crystals makes the nucleation of additional seedless $MoS_2$ flakes difficult. Therefore, if CVD conditions could be controlled to deposit fewer $MoO_xS_y$ particles onto the substrate, the formation of larger $MoS_2$ flakes will be promoted. While earlier works reported that the seeded growth mechanism mainly produces multilayer TMD flakes[37,63], $MoS_2$ obtained by our setup was dominated by the monolayer structure, with a much smaller second layer occasionally observed on some crystals. This is likely because the limited $MoO_3$ supply used in the experiments led to rapid reduction of the precursor supersaturation within the reactor, preventing the nucleation and growth of additional $MoS_2$ layers.

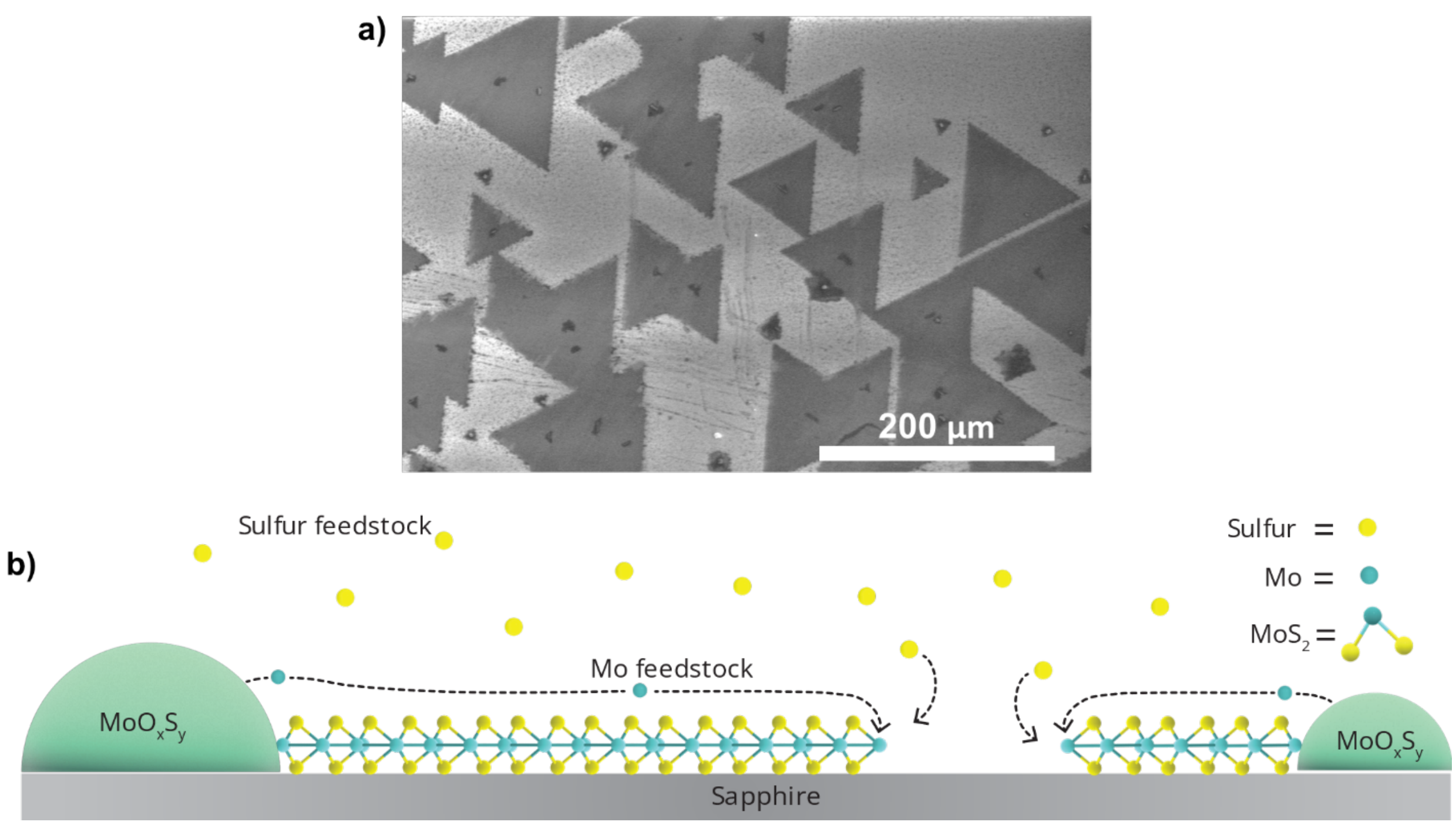

**Figure 7.** Seeded $MoS_2$ monolayer growth. (a) SEM image of grown $MoS_2$ flakes, with a nanoparticle visible at the center of each crystal. (b) Schematic of the seeded $MoS_2$ growth process.

## Conclusions

In this work, we established a robust framework for high-throughput analysis of the CVD growth of a large number of $MoS_2$ crystals by integrating in-situ optical microscopy with a semi-automated image processing pipeline. By tracking the morphological evolution of over 400 individual $MoS_2$ crystals throughout the CVD process, we extract quantitative, statistical insights into the underlying nucleation and growth kinetics. Our analysis reveals several important features of $MoS_2$ growth at high temperatures. We demonstrate that the $MoS_2$ growth rate is edge-attachment-limited, with precursor incorporation at the crystal periphery as the rate-determining step. Furthermore, $MoS_2$ crystals were found to grow in a non-competitive manner, in which neighboring $MoS_2$ do not compete for precursors because each crystal relies on its own $MoS_xO_y$ seed at the center to supply the Mo feedstock. While the nucleation rate of $MoS_2$ is sensitive to its van der Waals interaction with the sapphire substrate, its growth rate does not exhibit statistically significant dependence on the crystal/substrate orientation relation. These findings resolve existing ambiguities in the $MoS_2$ growth mechanism and provide useful guidance for selecting CVD process parameters to achieve desired material morphologies. In particular, our study shows that it could be a fruitful approach to constrain the deposition of $MoS_xO_y$ nanoparticles on the substrate to facilitate the growth of large $MoS_2$ crystals. The workflow developed here for analyzing in-situ observations is not limited to $MoS_2$ and serve as a versatile platform for investigating the growth dynamics of a wide range of 2D material systems. This methodology provides a valuable tool towards establishing predictive control over the scalable synthesis of 2D materials.

## Methods

### Miniaturized CVD

The miniaturized CVD system incorporates two distinct heating zones to separately control the evaporation rates of sulfur and $MoO_3$ precursors. $MoO_3$ was placed directly beneath the substrate. The optical microscope consists of an objective lens with ultra-long working distance and 10× magnification and a digital camera connected to it. It images the substrate through a dedicated window on the top of the furnace, generating a field of view of 1920x1080 pixels (or 1422 µm x 800 µm). The transparent sapphire substrate makes it possible for the optical microscope to image $MoS_2$ deposited on the bottom side of the substrate. A scratch was created on the substrate surface as a marker to aid microscope focusing (Figure S8). Experiments were recorded at 30 frames per second. Additionally, temperature profiles for both the sulfur and $MoO_3$ heating zones were recorded and synchronized with the video.

## Crystal Tracking Algorithm

A semi-automated image analysis workflow was developed to extract the growth history of each individual $MoS_2$ crystal appearing in the field of view from the video footage in a high-throughput manner. As illustrated in Figure S9, the workflow consists the following three modules:

### *1. Image pre-processing*

This module performs essential pre-processing of raw video data before tracking the evolution of each individual crystal. Because the $MoS_2$ growth rate shows a significant slowdown in the experiments, the original video footage is manually separated into two clips, corresponding to the fast vs slow growth periods. The first clip is resampled at a relatively high frequency of 5 Hz, and the second clip is resampled at 2 Hz. Resampling reduces the number of video frames for analysis by 7-9 folds. The original image frames are in RGB color. They are converted to gray scale by taking the weighted average of the red, green and blue channel intensities, yielding a single luminance value between 0 and 255. After cropping the frames, the region of interest (ROI) typically contains 600 – 2000 pixels in both dimensions, with a pixel size of $1.35 \times 1.35\ \mu m^2$. The image frames are then subject to a series of operations as described below.

*Histogram equalization* is applied to images to enhance the contrast between $MoS_2$ and the surrounding substrate. As the field of view is sometimes not evenly illuminated, the contrast-limited adaptive histogram equalization (CLAHE) algorithm is employed to remove the non-uniform illumination (Figure S10).

*Artifact removal*: pre-existing features on the substrate (e.g. dirt, surface defects) could be mistakenly identified as $MoS_2$ crystals in the segmentation step. Background subtraction is performed to remove these artifacts from images. This is accomplished by subtracting a frame containing only the initial bare substrate from the frames to be analyzed (Figure S11). We found that this simple procedure effectively eliminates stray artifacts and retrain only real crystals in the output images.

*Stabilization* of image sequence is implemented in commercial software (Topaz Video AI) to reduce video shaking or drifting due to the vibration of the optical microscope stage. This operation improves the accuracy of tracking crystals over time, which relies on the centroid locations to identify the same crystals in different frames.

*Segmentation*: processed image frames are binarized by histogram thresholding to distinguish $MoS_2$ crystals from the substrate. We found that traditional thresholding algorithms such as the Otsu's method do not yield consistent segmentation results across different frames. As a result, manual threshold adjustment needs to be applied to many frames, rendering the procedure extremely cumbersome. To address this issue, a supervised machine learning model based on a pixelwise random forest classifier was developed for the segmentation task. To train the model, only one frame from the video footage was manually labeled to generate the training dataset. This frame was selected based on the criterion that it provides a balanced representation of the $MoS_2$ crystal and substrate regions. It was then divided into 52 smaller patches of 256 × 256 pixels with 50% overlap between adjacent regions. As shown in Figure S12, each patch in the training and test sets was passed through 12 image filters, including Gabor (0°, 45°, 90°, and 135°), Roberts, Sobel, Prewitt, Scharr, Gaussian, median, and variance filters. These filters are commonly used for edge detection, texture characterization, and feature extraction. Along with the original grayscale image, they generate 13 feature descriptors for each pixel, which serve as input to a random forest (RF) model with 100 trees. The RF model was trained to classify pixels as crystal (1) or background (0), which achieves a 96% accuracy on the validation set (520 256×256 images). Figure S13 shows an example of a video frame segmented by the RF model. It only took 3-4 hours to prepare the training and validation sets, less than 30 minutes to train the RF model and around 1 hour to binarize all the frames in a video, which is substantially faster than manual thresholding.

***2. Crystal tracking***

In this module, each crystal that appears in the footage is assigned a unique identification (ID) number. This ID number is the same for a specific crystal across different frames so that its shape evolution could be tracked. A binarized image sequence is analyzed frame-by-frame in the forward direction. For each frame, isolated objects in the ROI are first determined by the contour detection method. To avoid noise being mistakenly recognized as a crystal, an object is labeled a crystal cluster only when its area exceeds a minimum size of 15 pixels (or ~27 $\mu m^2$). As such, the nominal nucleation time of a crystal detected by the algorithm differs from the real nucleation time due to the resolution limitation of the optical microscope and the image analysis method, although they are correlated. Objects detected in a frame will first be compared with crystals present in the previous frame. If the centroid of an object is within a threshold distance of 15 pixels from an already identified crystal, they are deemed the same crystal and each pixel in the object is labeled by the crystal ID. Otherwise it is treated as a new $MoS_2$ nucleus and assigned a new ID.

Several complications could arise in this module and are addressed as follows. First, due to segmentation errors, some small crystals may disappear from the ROI after being detected in a previous frame, which is unphysical. A "redraw" step is introduced to correct this type of error. If a crystal is detected at a given location in an early frame but disappears subsequently, the crystal will be "copied" from the previous frame and "pasted" onto the same location with the same ID.

Second, crystals near the ROI borders could grow out of the field of view in the footage, rendering their shape evolution not completely known. These crystals are excluded from the analysis to prevent erroneous statistics. Therefore, whenever the tracking algorithm detects a crystal object touching a border, it filters out the crystal from the binarized image frames (Figure S14) and removes it from the tracked crystal list.

Lastly, when a crystal merges with its neighbor(s), the coalesced polycrystal could potentially be misidentified as a single crystal object, leading to incorrect tracking. It is thus necessary for the workflow to distinguish between single-crystalline and polycrystalline objects and analyze them separately. We employed supervised learning techniques to classify the two types of crystalline objects based on their shape descriptors. The descriptors used for classification include circularity, outer/inner circle circumference, inner/outer regular triangle circumference, and shape invariants values (Hu moments). The Hu moments are a set of seven numerical features derived from image moments that are invariant to translation, rotation, and scaling, making them particularly effective for distinguishing geometric shapes. The training dataset consists of ~1000 objects manually labeled as single- or poly-crystals, which was augmented to a total of 2340 objects. Figure S15a compares the performance of three different classifiers: RF, k-nearest neighbors (kNN) and support vector machine (SVM). For the kNN classifier, the number of neighbors was set to 50 with Euclidean distance (Minkowski, $p = 2$) and uniform weighting. The SVM classifier employs a radial basis function (RBF). The RF classifier was trained with 100 decision trees using the Gini impurity criterion, unrestricted maximum depth, and a fixed random state for reproducibility. When applied to a validation dataset containing ~1560 crystal objects, kNN, SVM and RF achieve prediction accuracies of 65.1%, 97.3% and 99.7%, respectively. The RF model outperforms the other classifiers and is adopted in the tracking module. When a polycrystalline object is identified in a frame, the algorithm searches the image sequence backward to determine the IDs of individual single crystals it contains. The polycrystal is labeled by a tuple of these ID numbers, which allows the coalescence process to be reconstructed. Eventually, two image sequences result from the tracking process: one contains indexed single crystals only and the other records distinct polycrystalline clusters (Figure S15b).

### *3. Crystal feature extraction*

After each unique $MoS_2$ crystal is identified and labeled throughout the image sequence, the workflow extracts time-dependent shape features of all the crystals from the labeled frames. These features include centroid location, orientation, size and boundary contour.

Because segmented crystals are often not perfectly polygonal, we first fit their images with equilateral triangles to extract the orientations of single-crystalline $MoS_2$ (Figure S16a). Specifically, the area and centroid location of a crystal are first determined from the object image. A regular triangle is initialized with the same area and center as the measured values and a randomly assigned orientation. The orientation ($\theta$) and the side length of the triangle are then iteratively optimized to maximize the intersection over union (IoU) between the regular triangle and the crystal image. After convergence is reached, the returned $\theta$ value is used as the crystal orientation. The differential evolution algorithm is used for optimization as it efficiently explores continuous parameter spaces without requiring an exhaustive search. Polycrystalline $MoS_2$ flakes are fitted by two or more regular triangles in a similar way (Figure S16b). The centers of the triangles are placed at the centroids of the individual constituent crystals identified from previous frames.

The final outputs of the entire workflow include a list of unique $MoS_2$ single- and poly-crystals, their shape features as a function of time, and records of crystal coalescence events (time and participating crystal IDs). Such information constitutes a complete digitized representation of the in-situ observation, from which the $MoS_2$ nucleation and growth process could be reconstructed and analyzed.

**Acknowledgement**

FA gratefully acknowledges support from the Fulbright Fellowship Program. ALH was supported by the Air Force Office of Scientific Research under Award No. FA9550-21-1-0460. MT was supported by the Air Force Office of Scientific Research under Award No. FA9550-23-1-0444. TZ, YW, JZ, and JL acknowledge the support by the NSF I/UCRC Center for Atomically Thin Multifunctional Coating (ATOMIC) under award no. EEC-213882, NSF PFI grant no. 2431965, and the Welch grant C-2248.

**Author Contributions**

MT and JL conceived and supervised the project. TZ, YW, JZ and JL designed and performed the experiments. FA, ALH and MT designed the in-situ video processing workflow and analyzed the data. FA, TZ, JL and MT drafted the original manuscript. All authors discussed the results and revised the manuscript.

**Competing interests**

The authors declare no competing interests.

**Supplementary Information for**

# Revealing the $MoS_2$ Growth Mechanism in Chemical Vapor Deposition: Real-Time Imaging and Statistical Analysis

Faizal Arifurrahman[1,2†], Tianshu Zhai[1†], Yuguo Wang[1], Jing Zhang[1], Andrew L. Hitt[1], Jun Lou[1,3*], Ming Tang[1,3*]

1. Department of Materials Science and Nanoengineering, Rice University, Houston, TX

2. Faculty of Mechanical and Aerospace Engineering, Institut Teknologi Bandung, Jalan Ganesha 10, Bandung, 40132, Indonesia

3. Rice Advanced Materials Institute, Rice University, Houston, TX 77005, USA

[†] These authors contributed equally.

[*] Corresponding author emails: jlou@rice.edu, mingtang@rice.edu

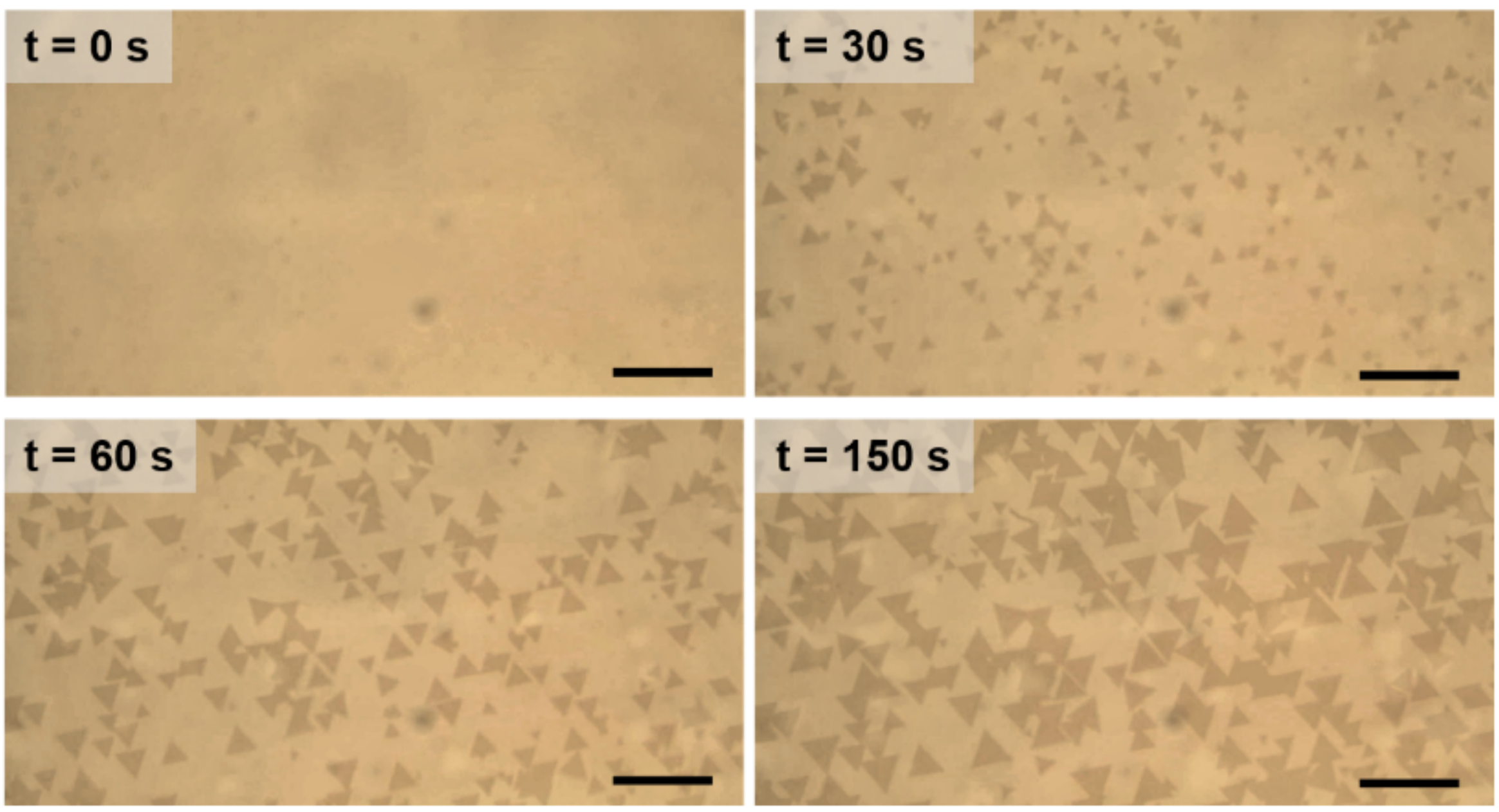

**Figure S1.** Snapshots of video footage from in-situ optical microscopy at $t$ = 0, 30, 60 and 150 s after $MoS_2$ crystals were first detected on the substrate. Scale bar is 200 μm.

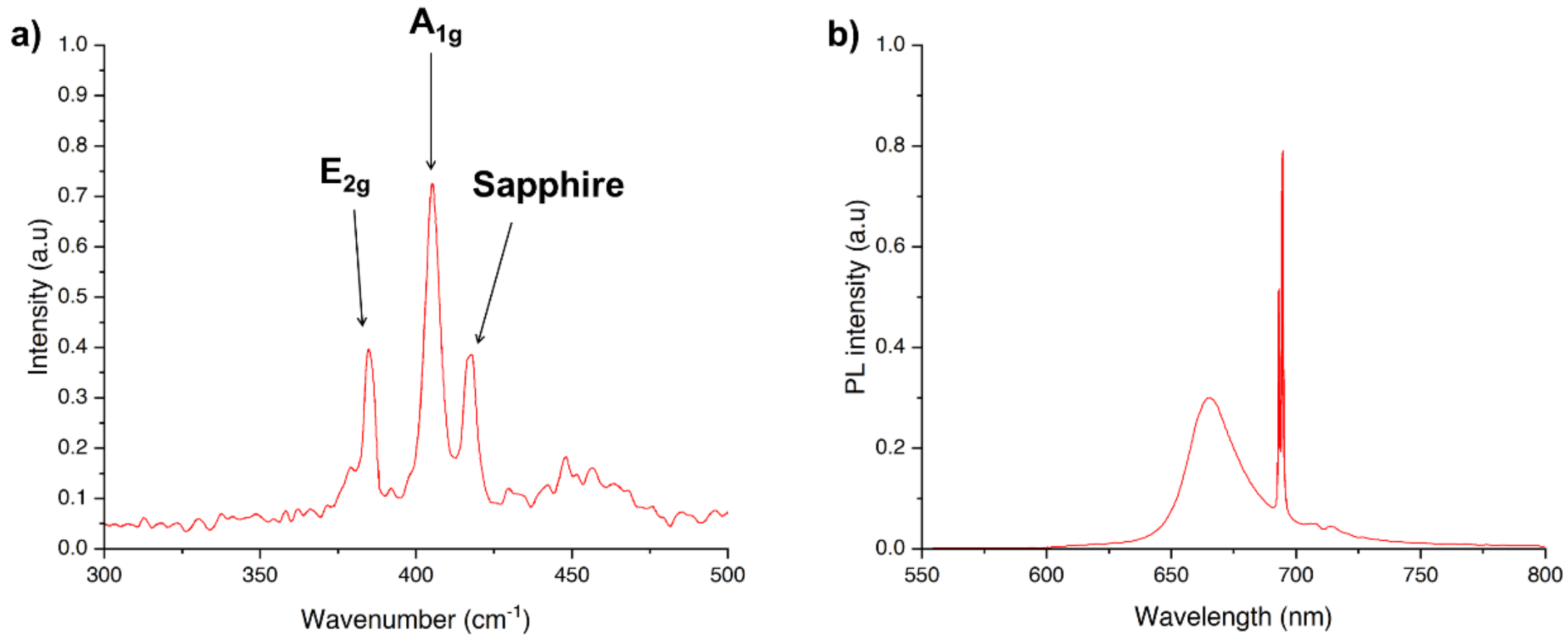

**Figure S2.** Raman and Photoluminescence (PL) spectra of monolayer MoS2 grown on the sapphire substrate in the miniature CVD system. (a) Raman and (b) PL spectra.

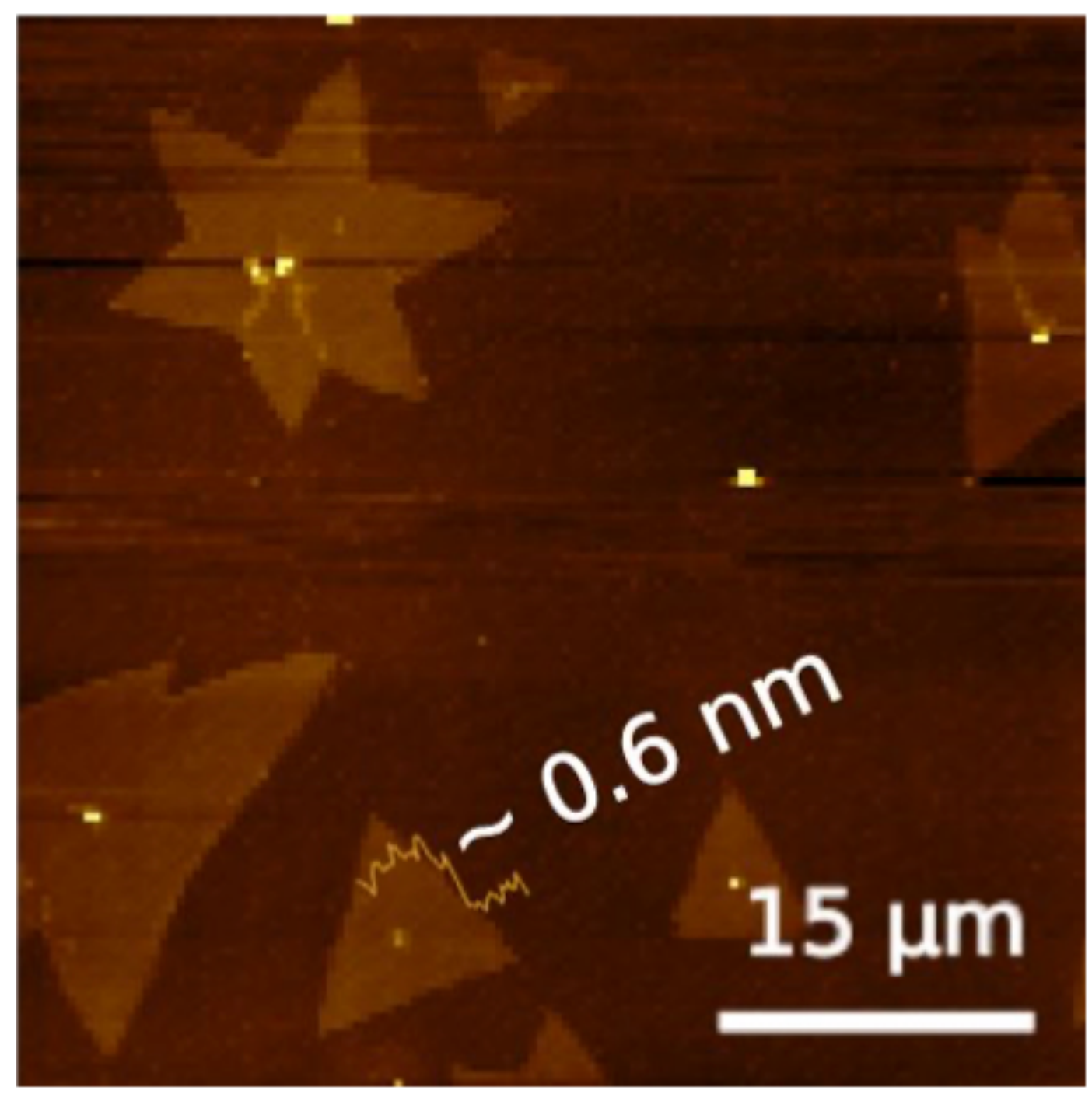

**Figure S3.** AFM height image of monolayer $MoS_2$ grown on sapphire, superimposed with the height profile across a flake edge.

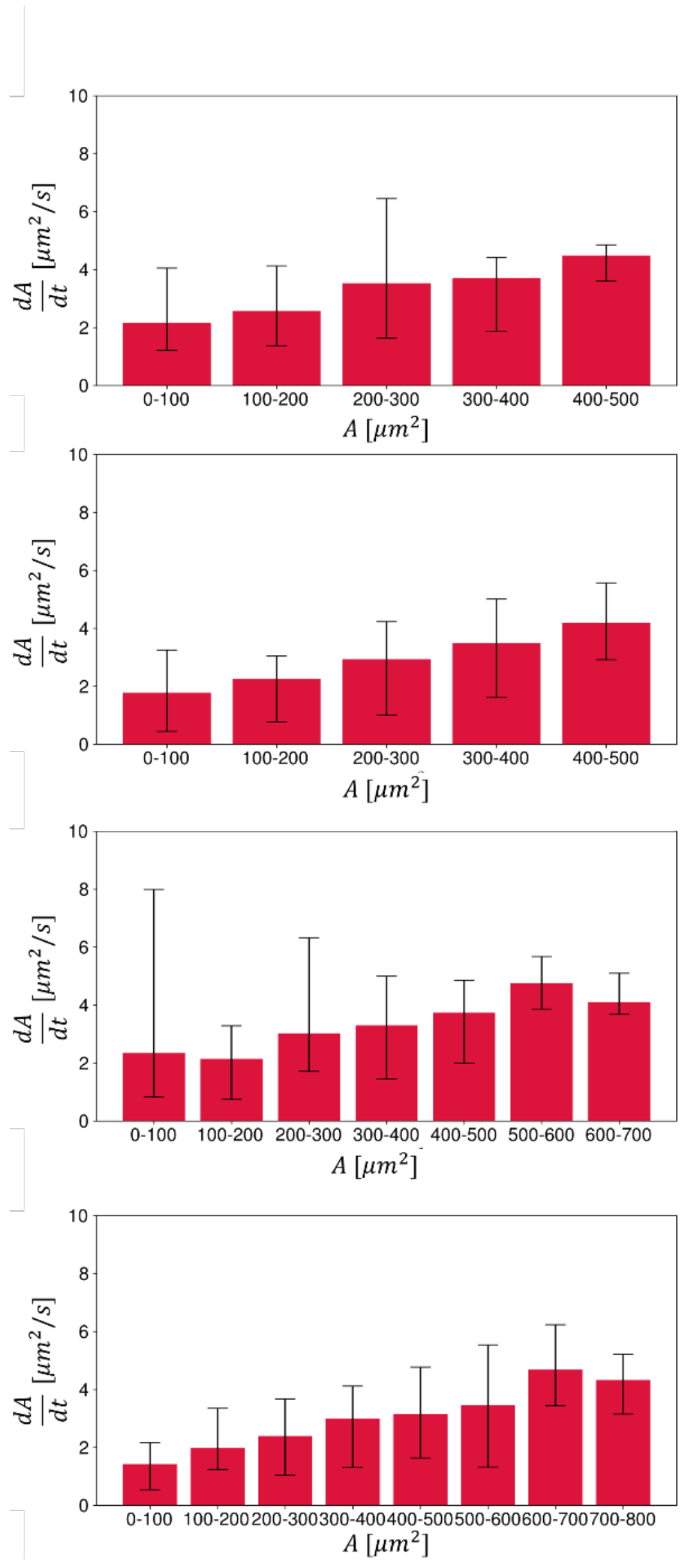

**Figure S4.** Instantaneous $MoS_2$ growth rate $dA/dt$ versus crystal area at a) 25, b) 30, c) 35, and d) 40 s after the first $MoS_2$ was detected.

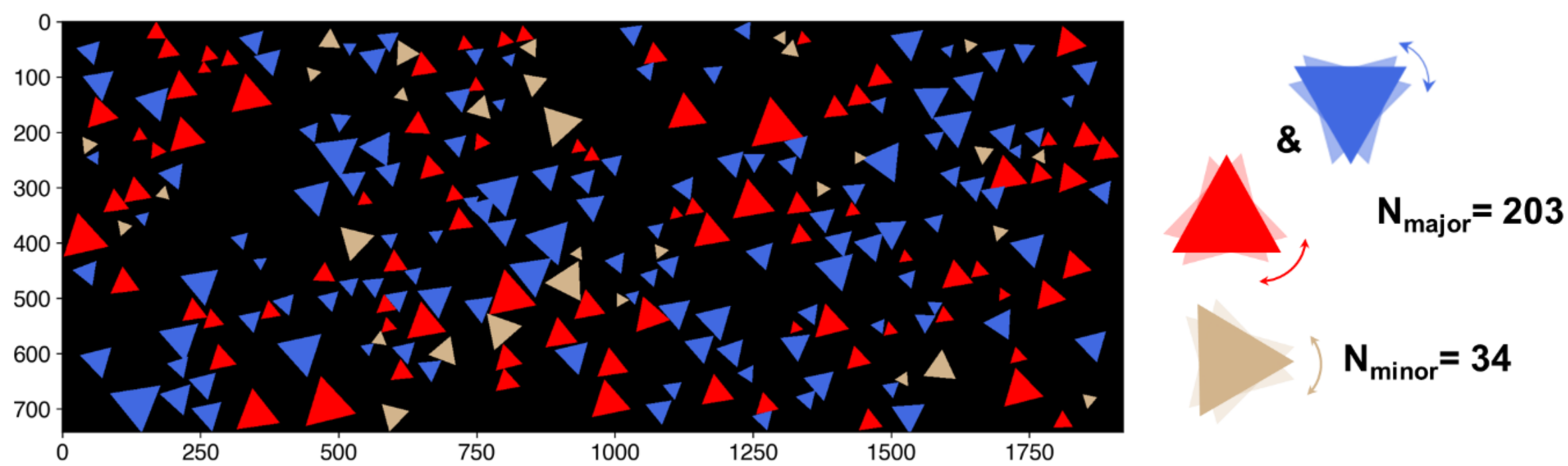

**Figure S5.** Spatial distribution of $MoS_2$ crystals with different orientations: blue – 0°, red – 60°, and tan – other. A 15° tolerance is used when assigning crystals to the 0° or 60° orientation group.

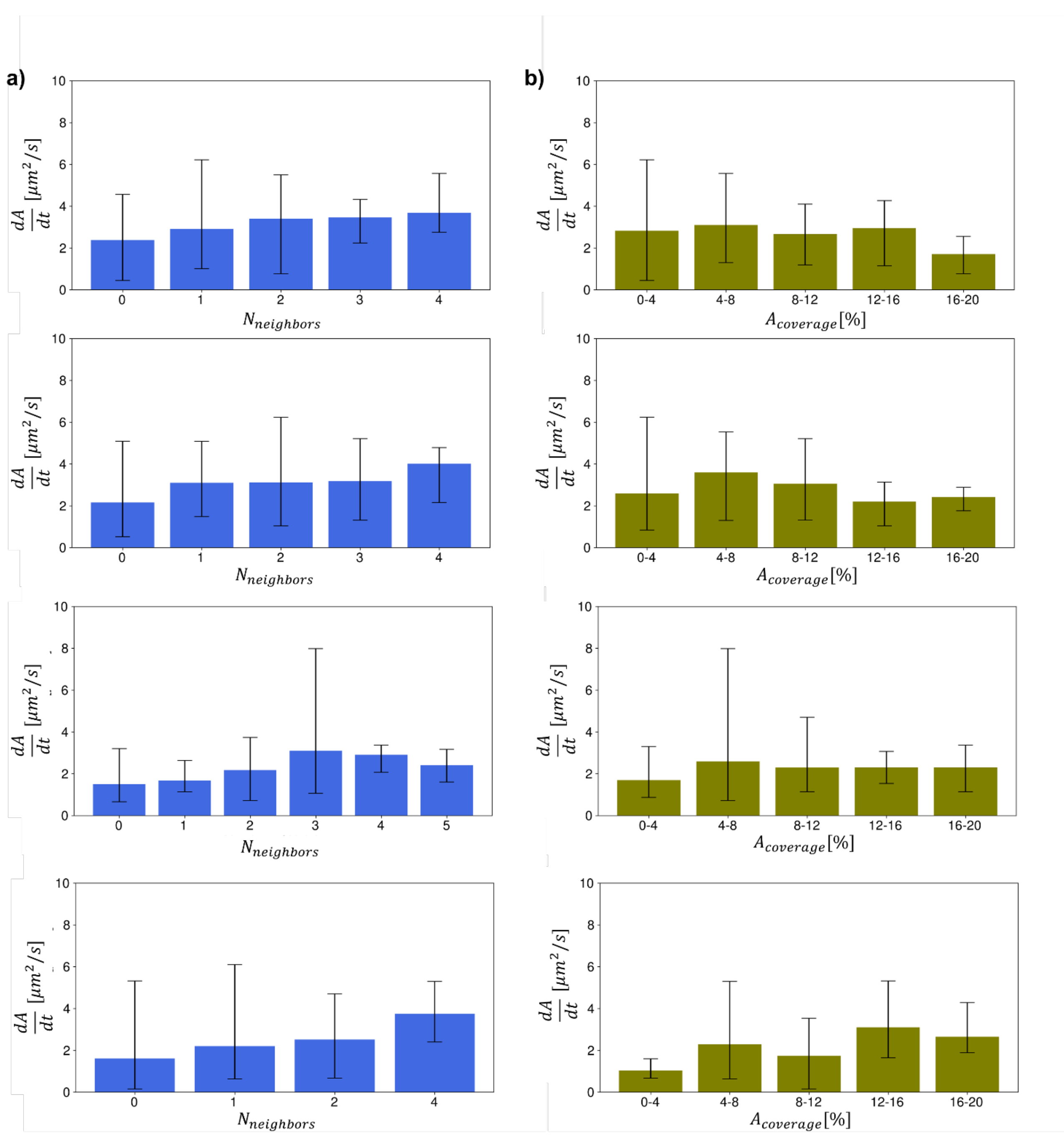

**Figure S6.** Instantaneous crystal growth rate vs a) number and b) area coverage of neighbor crystals at 30, 40, 60, and 70 s after the first $MoS_2$ crystal was detected.

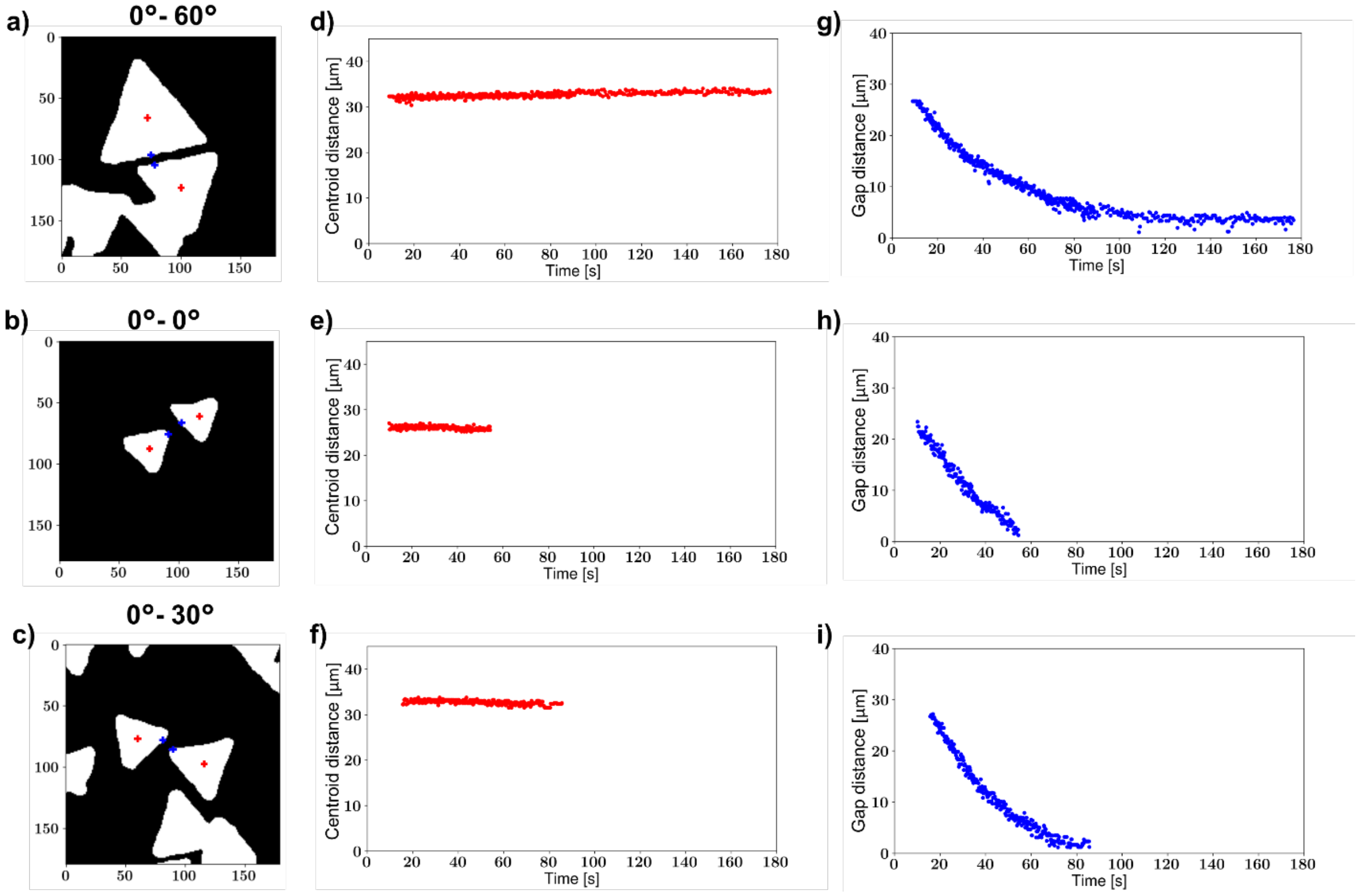

**Figure S7.** a-c) $MoS_2$ crystal pairs with different orientation relations selected from the in-situ video footage. d-e) Centroid distance of the crystal pair as a function of time. g-i) Gap distance of the crystal pair as a function of time.

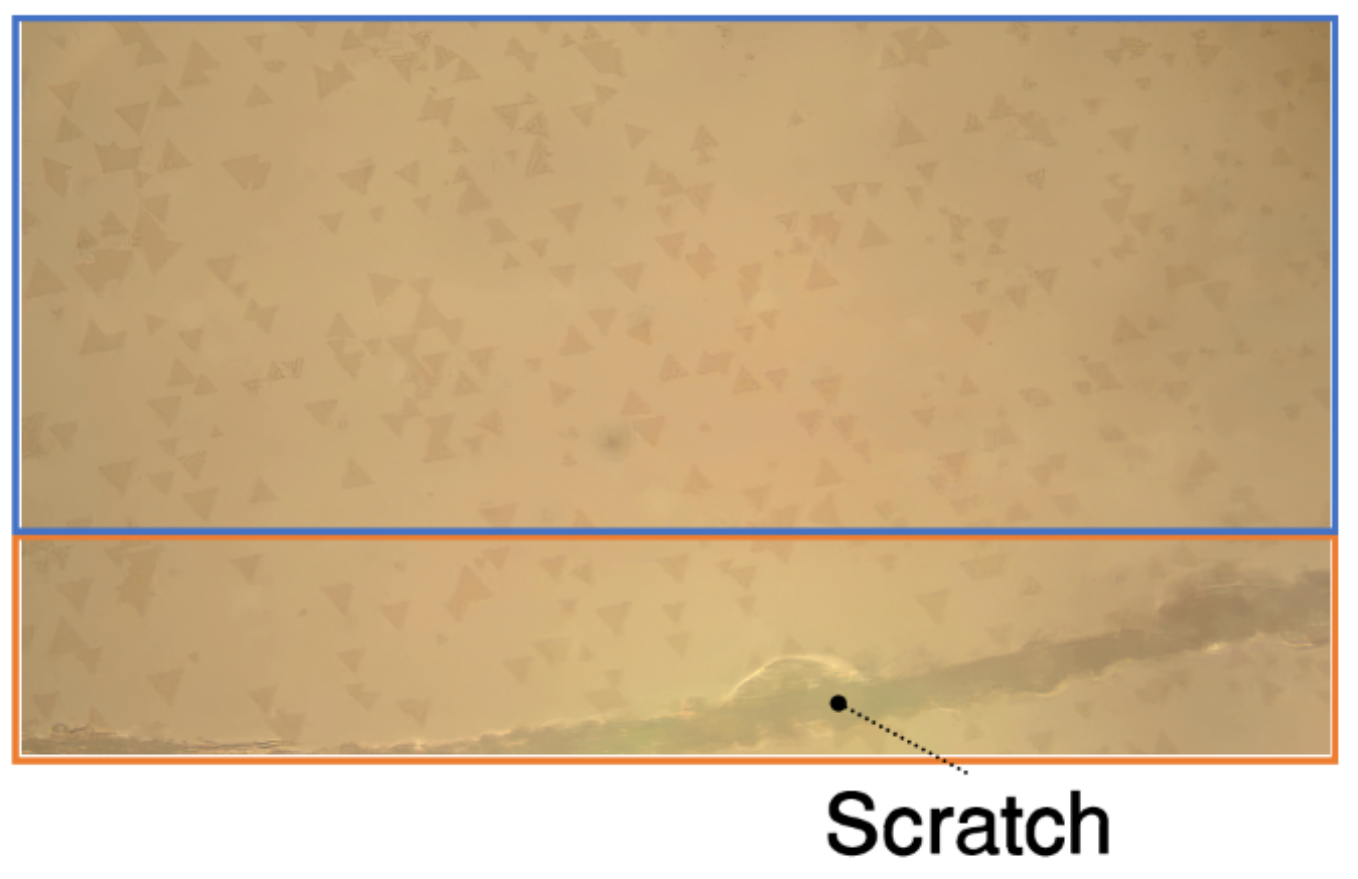

**Figure S8**. Substrate was intentionally scratched to help focusing the optical microscope on the substrate surface.

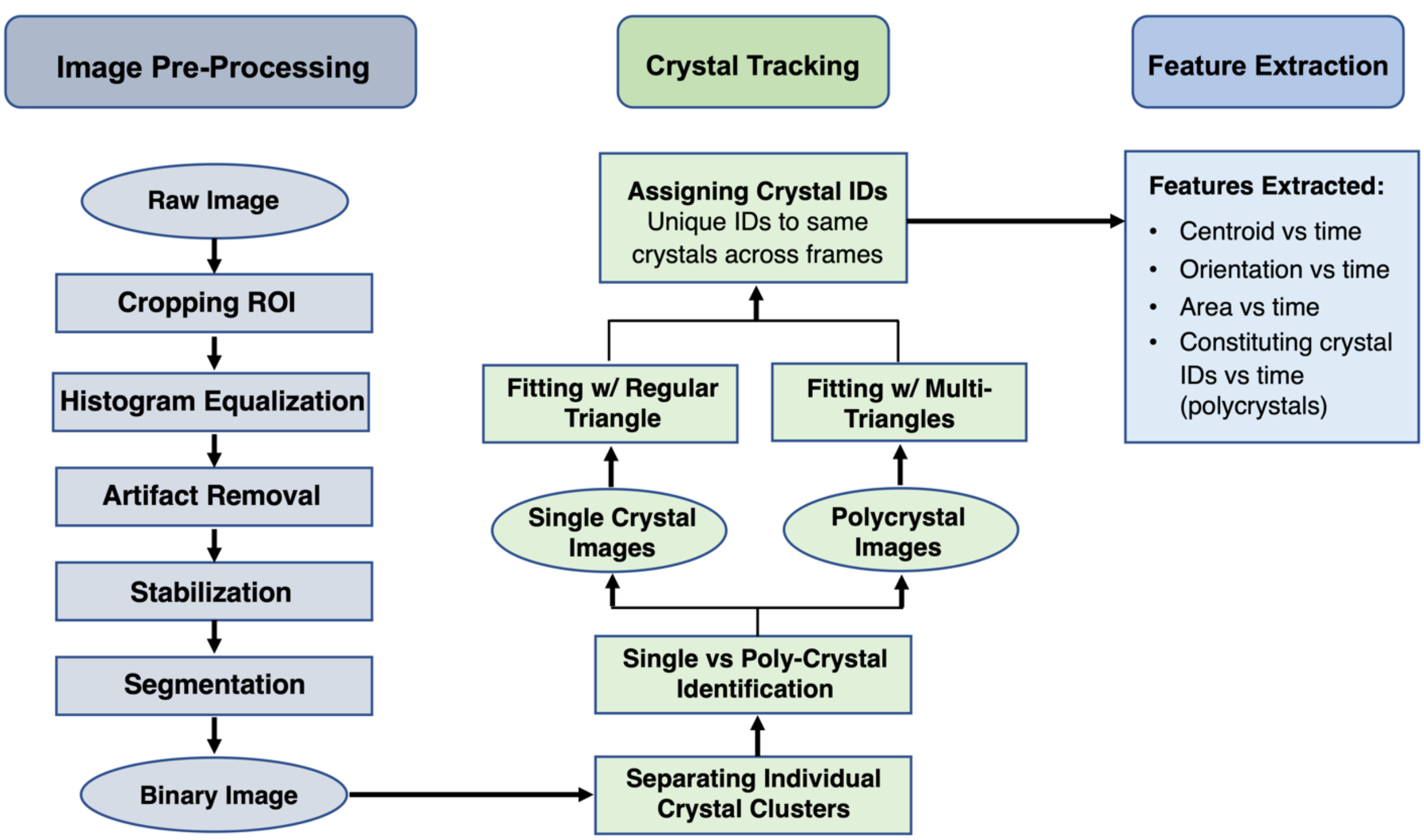

**Figure S9.** Structure of the crystal tracking algorithm.

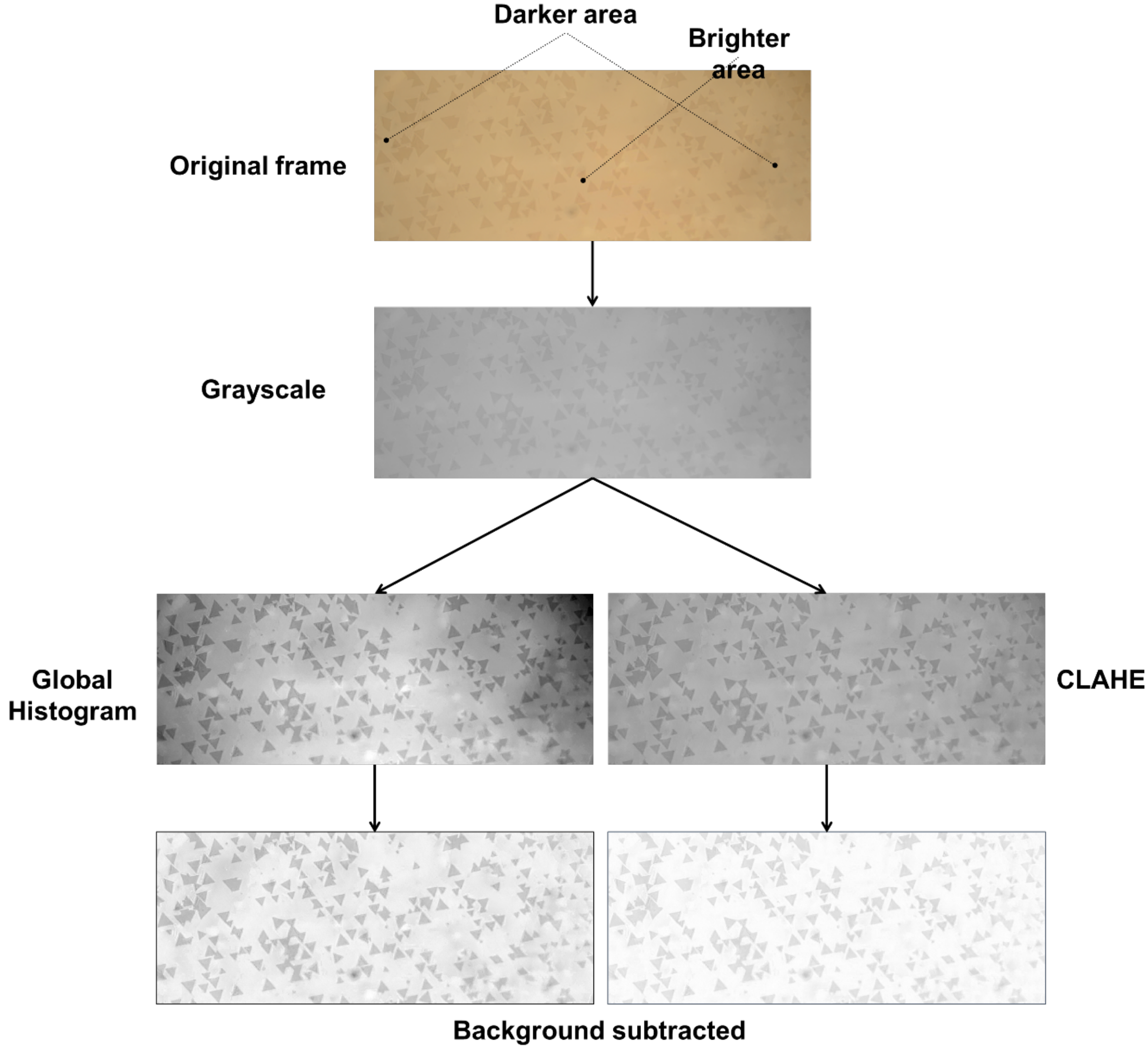

**Figure S10.** Image pre-processing: application of the CLAHE algorithm to enhance $MoS_2$ / substrate contrast. Compared to global histogram equalization (left), CLAHE reduces illumination variation across the images and enables more effective separation of crystals from the substrate in subsequent steps.

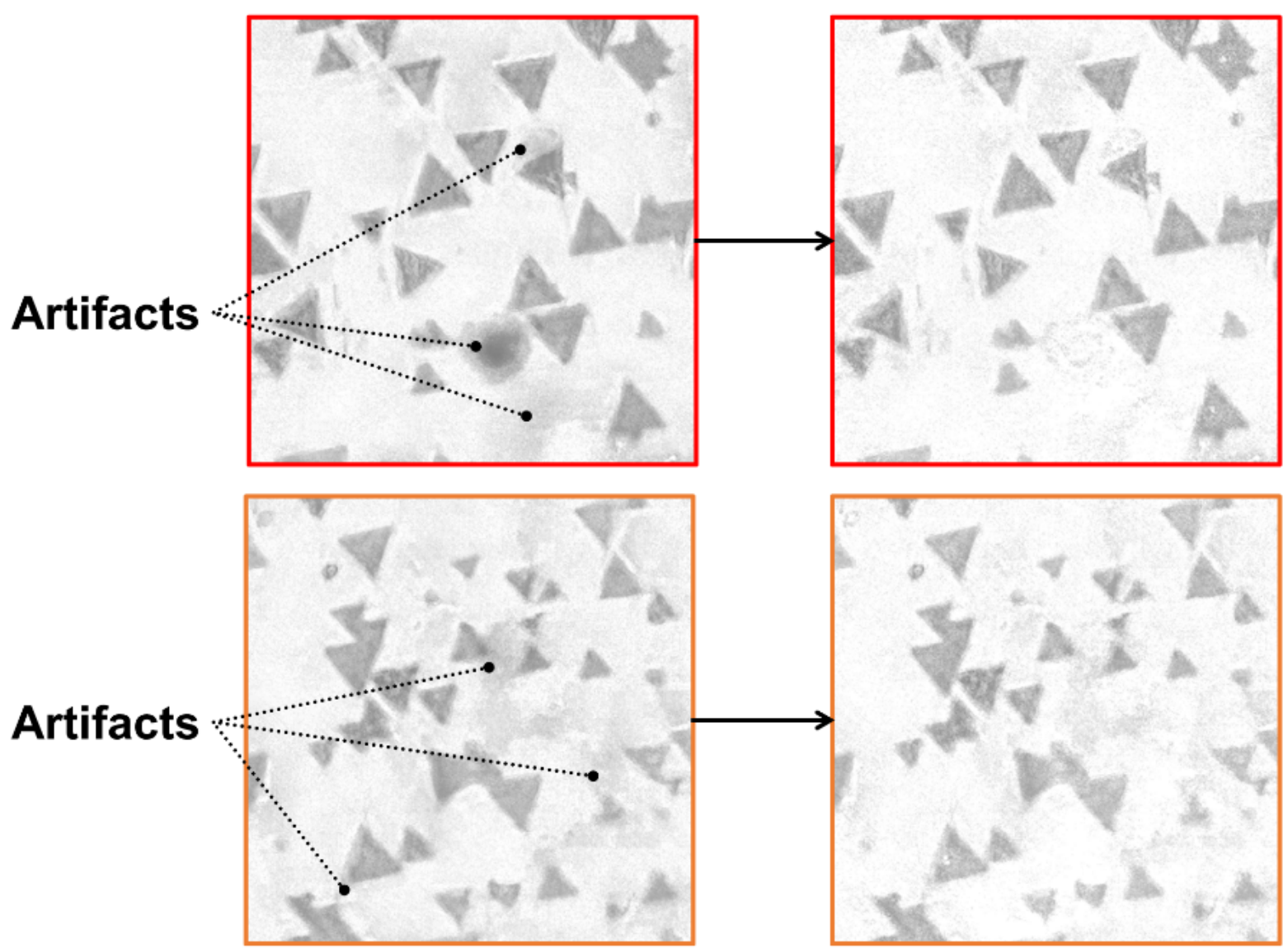

**Figure S11**. Image pre-processing: remove pre-existing artifacts (e.g. stain, dirt, etc.) by subtracting an early frame containing only the bare substrate from all subsequent frames. This operation prevents artifacts from being mis-identified as $MoS_2$ crystals.

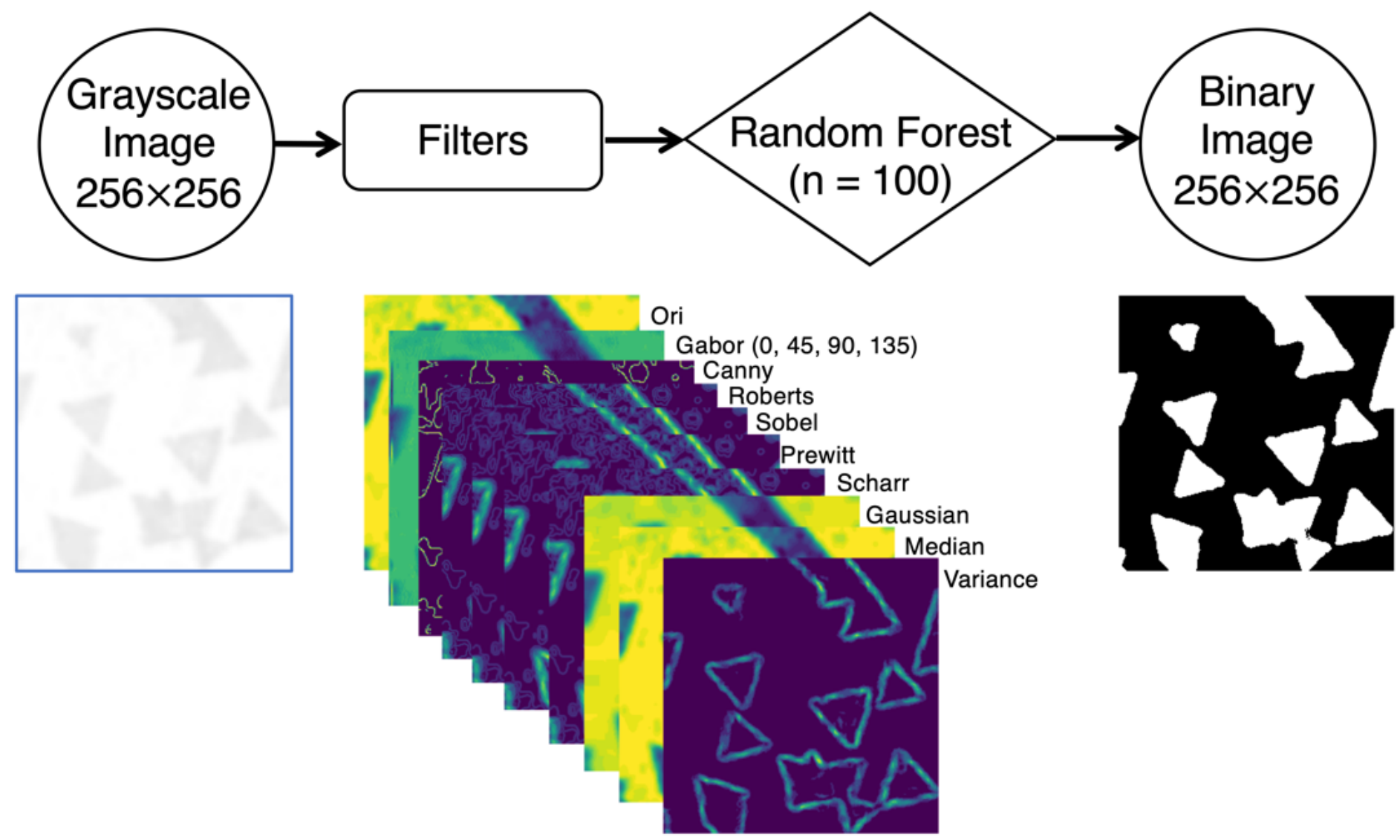

**Figure S12.** Random forest model for segmenting crystals (white) and substrate (black) in video frames.

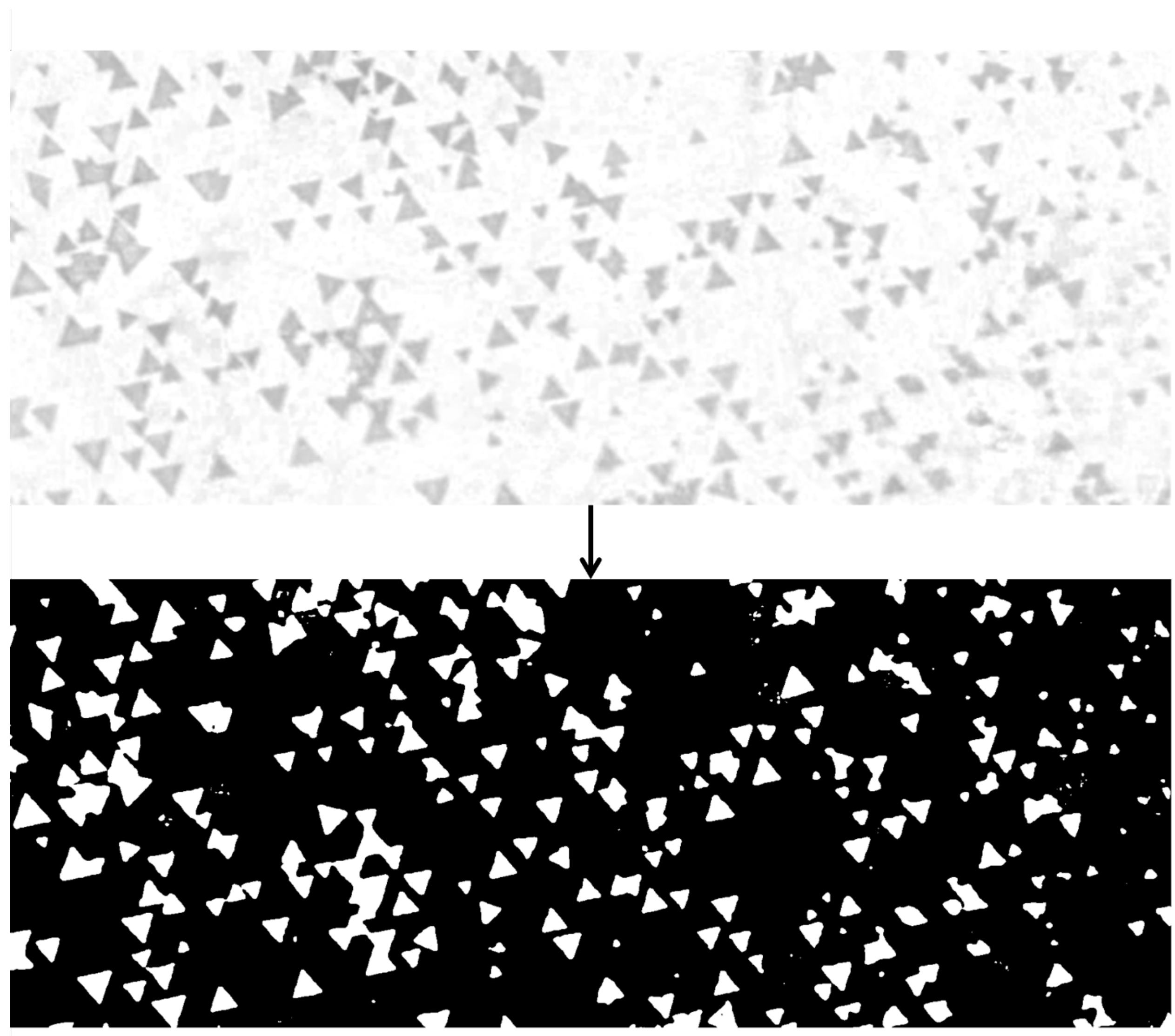

**Figure S13.** Example of a binary image (bottom) segmented from the grayscale image (top) by the random forest model.

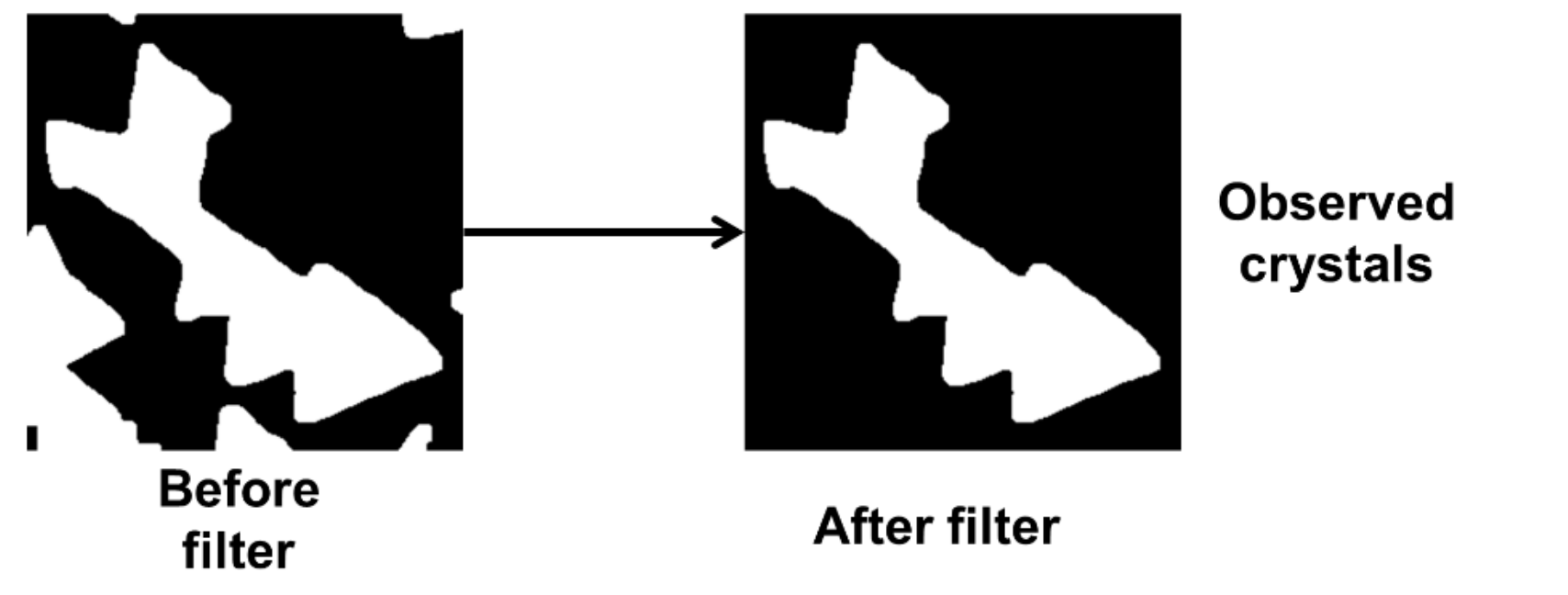

**Figure S14.** Example of a crystal cluster that touches the image border being excluded from analysis in the crystal tracking module.

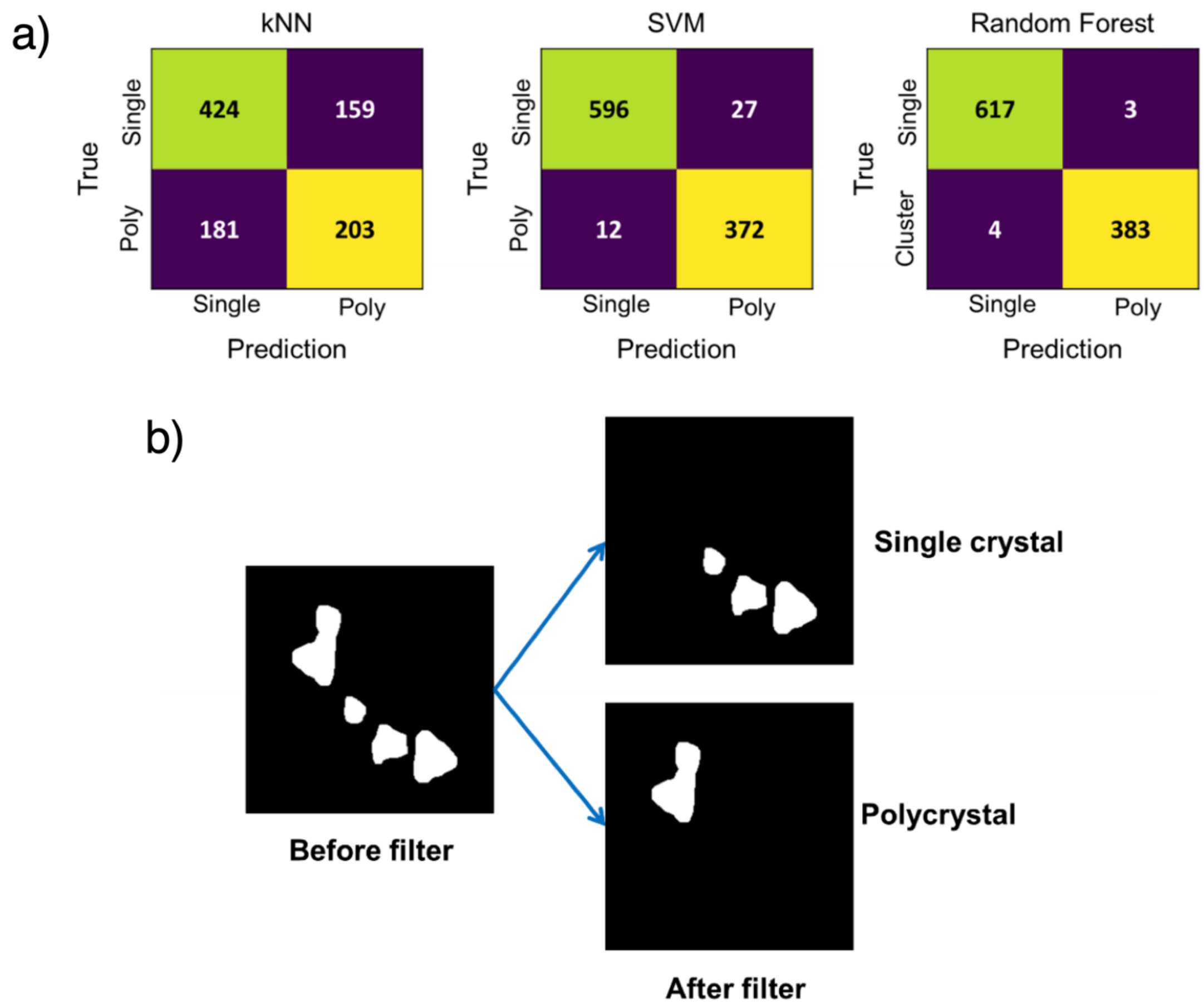

**Figure S15.** (a) Confusion matrix of kNN, SVM and Random Forest classifiers when applied to detect whether a crystal cluster is a single crystal or polycrystal. (b) Polycrystalline and single-crystalline $MoS_2$ flakes present in a frame are classified and analyzed separately in the tracking process.

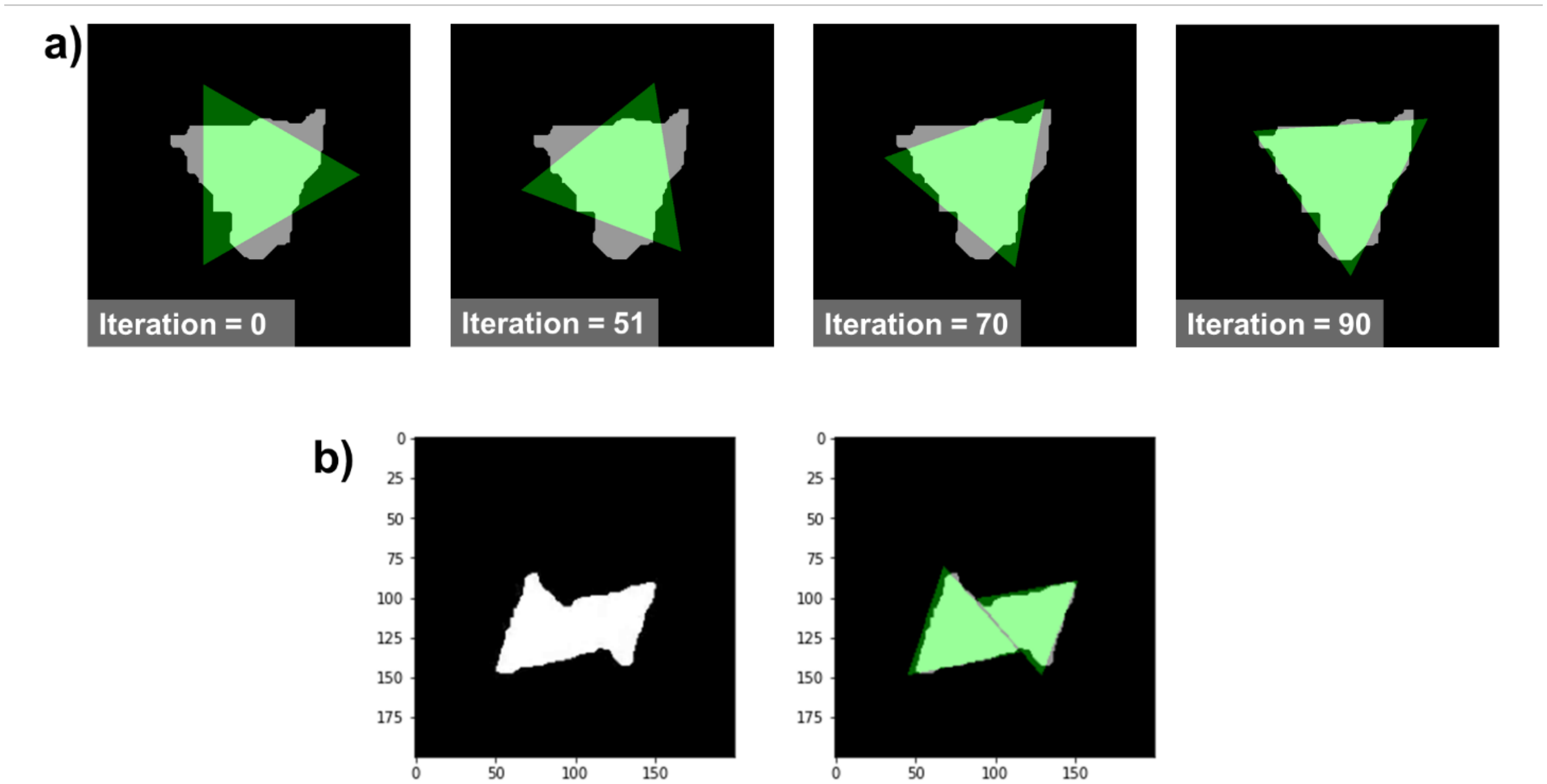

**Figure S16.** a) Fitting a) a $MoS_2$ single crystal, and b) a $MoS_2$ polycrystal with regular triangle(s) using the differential evolution algorithm.